\documentclass{elsart}

\usepackage{units}

\usepackage{color}

\usepackage{amssymb}
\usepackage{amsmath}
\usepackage[dvips]{graphicx}
\usepackage{psfrag}
\usepackage{floatflt}
\usepackage{overpic}
\usepackage{lineno}

\newcommand{\md}{\mathrm{d}}

\journal{Astroparticle Physics}

\begin{document}

\begin{frontmatter}



\title{REAS3: Monte Carlo simulations of radio emission from cosmic ray air showers using an ``end-point'' formalism}


\author[UNIIEKP]{M. Ludwig\corauthref{cor}},
\corauth[cor]{Corresponding author.}
\ead{marianne.ludwig@kit.edu}
\author[FZKIK]{T. Huege}

\address[UNIIEKP]{Karlsruher Institut f\"ur Technologie, Institut f\"ur Experimentelle Kernphysik - Campus S\"ud, 76128 Karlsruhe, Germany}
\address[FZKIK]{Karlsruher Institut f\"ur Technologie, Institut f\"ur Kernphysik - Campus Nord, Postfach 3640, 76021 Karlsruhe, Germany}

\begin{abstract}
	
	In recent years, the freely available Monte Carlo code REAS for modelling radio emission from cosmic ray air showers has evolved to
	include the full complexity of air shower physics.
	However, it turned out that in REAS2 and all other time-domain models which calculate the radio emission by superposing the radiation of the
	single air shower electrons and positrons, the calculation of the emission contributions was not fully consistent. In this article, we present a 
	revised implementation in REAS3, which incorporates the missing radio emission due to the variation of the number of
	charged particles during the air shower evolution using an ``end-point formalism''.
	With the inclusion of these emission contributions, the structure of the simulated radio pulses changes
	from unipolar to bipolar, and the azimuthal emission pattern becomes nearly symmetric. Remaining asymmetries can be explained
	by radio emission due to the variation of the net charge excess in air showers, which is automatically taken into account
	in the new implementation. 
	REAS3 constitutes the first self-consistent time-domain implementation based on single particle emission taking the full complexity
	of air shower physics into account, and is freely available for all interested users.
	
\end{abstract}

\begin{keyword}
cosmic rays \sep extensive air showers \sep electromagnetic radiation from moving charges \sep computer modeling and simulation
\sep radio emission \sep endpoint formalism
\end{keyword}

\end{frontmatter}


\section{Introduction}\label{ch:introduction}

	High energy cosmic rays initiate extensive air showers when entering the Earth's atmosphere. The electromagnetic component of the air
	shower produces radio emission which can be measured to study the characteristics of the cosmic rays. Radio detector arrays like 
	LOPES \cite{Falcke05}, \cite{HuegeArena2010} and CODALEMA 	\cite{Ardouin05}, \cite{Ardouin09} have verified that radio emission in the air
	is dominated by a geomagnetic effect, and they have studied correlations of the radio signal with shower parameters in great
	detail. With AERA 
	\cite{Berg09} within the Pierre Auger Observatory \cite{Abra04} and LOFAR \cite{LOFAR03}, radio detection will be applied
	on larger scales and new ``super-hybrid'' techniques for measuring cosmic ray air showers will be employed. To understand the 
	measurements and to learn more about the physics of cosmic rays using radio signals from air showers, a solid
	theoretical understanding of the radio emission 
	process is needed. Presently, two major approaches exist, both of which are mainly based on geomagnetic effects \cite{HuegeArena2008}. \\
	On the one hand, there is a model based on the calculation of emission from the deflection of single particles in the geomagnetic field which was mainly investigated by Huege et al. \cite{HuegeFalcke2003a},
	\cite{HuegeFalcke2005a}, \cite{HuegeFalcke2005b}, \cite{HuegeUlrichEngel2007a}, 
	in particular with the implementation in the REAS Monte Carlo code.
    The treatment of the radio emission with Monte Carlo techniques makes it straight-forward to couple it with detailed Monte Carlo air shower 	
    simulations. In case of REAS, this is done
    with CORSIKA \cite{Corsika} in which the needed particle distributions are histogrammed by COAST \cite{HuegeUlrichEngel2007a}. It is
    interesting to note that the original frequency-domain calculation of this implementation
	predicted spectra decaying to zero at small frequencies \cite{HuegeFalcke2003a}, the time-domain implementation in the 
	simulation code REAS, however, hitherto predicted
	spectra levelling off at a finite field strength for small frequencies, leading to essentially unipolar pulses 
	\cite{HuegeFalcke2005b}, \cite{HuegeUlrichEngel2007a}. \\
	On the other hand, the macroscopic description of geomagnetic radiation (MGMR) developed by Scholten et al.
	\cite{ScholtenWernerRusydi} constitutes a modern implementation of the approach for radio emission modelling by Kahn and Lerche in 1966 \cite{KahnLerche}. Due to
	the Lorentz force, moving electrons and positrons are separated in the Earth's magnetic field. In MGMR, this is described by a 
	time-variable net electric current in the electron-positron plasma, which is moving through the atmosphere with approximately 
	the speed of light. The emission of an electromagnetic pulse is caused by these time-dependent transverse currents. Due to
	the dependence on the evolution of the charged particles in the shower with time, the MGMR model predicts a bipolar structure 
	of the pulse shape \cite{ScholtenWernerRusydi}. \\	
	The differences in the results of both models were studied 
	and led to the conclusion that at least one model was not complete. The main reason for the different results of both approaches 
	was that in REAS2 
	(and all other time-domain approaches based on single particle emission, e.g. \cite{ReAIRES}, \cite{SuprunGorham2003}),
	emission due to the variation of the number of charged particles within the shower was not considered.
	A detailed discussion of this issue can be found in \cite{HuegeLudwigScholten} to which we kindly refer the reader for further details.
	To correct the implementation in REAS, these missing radiation contributions had to be taken into account.
	It should be stressed that the resulting implementation, based on the ``end- point'' formalism, consistently describes the radiation of the
	complete underlying particle motion, not just ``synchrotron''-like emission from
	the particle acceleration (see \cite{James2010} for a discussion of the universal nature of the end-point formalism). 

\section{Simulation algorithm of REAS}\label{ch:algorithm}

	From REAS2 to REAS3, the simulation algorithm is unchanged and the air shower information is provided by CORSIKA and COAST
	in the same way	as before. Therefore, we here only give a short overview of the technical implementation and for details, we kindly refer the 
	interested reader to \cite{HuegeUlrichEngel2007a}.\\
	First, the shower is simulated with CORSIKA using the air shower parameters of interest (such as primary energy,
	magnetic field, mass of primary, incoming direction, etc.). Using COAST, the information of the electrons and positrons is
	saved in histograms. These histograms contain information about the atmospheric depth of the particle, the particle arrival time,
	the	lateral distance of the particle from the shower axis, the particle energy and the particle momentum direction. In the next step, REAS is
	generating individual electrons and	positrons randomly according to the histogrammed distributions.	These particles are then tracked
	analytically in the geomagnetic field. Note that the trajectories of the REAS simulation do not represent real trajectories
	of particles, i.e. one long particle trajectory is represented by an ensemble of several shorter, unrelated trajectories in the code. The
	length of the trajectories is determined by a parameter $\lambda$ which is explained in chapter \ref{ch:crosschecks}. \\
	In REAS3, the trajectory is distributed symmetrically around the position at which the particle is generated.	
	
\section{Contributions due to charge variation}\label{ch:contributions}
	
	In REAS2 \cite{HuegeUlrichEngel2007a}, the radiation of single particles of an air shower is calculated as
	\begin{align}
	\vec{E}(\vec{x},t) = e \left[ \frac{\vec{n}-\vec{\beta}}{\gamma^2(1-\vec{\beta}\cdot\vec{n})^3R^2}\right]_{\mbox{ret}} +
	\frac{e}{c} \left[ \frac{ \vec{n}\times [(\vec{n}-\vec{\beta})\times \dot{\vec{\beta}}]} 
	{(1-\vec{\beta}\cdot\vec{n})^3 R}\right]_{\mbox{ret}} ,\label{eq:radEq}
	\end{align}
	where $e$ indicates the particle charge, $\vec{\beta} = \vec{v}(t)/c$ is given by the particle velocity, $R(t) = \vert\vec{R}(t)\vert$
	describes the vector between particle and observer position, $\vec{n}(t) = \vec{R}(t)/R(t)$ is the line-of-sight direction between 
	particle and observer, and $\gamma$ is the Lorentz factor of the particle. The index ``ret'' means that the equation needs to be
	evaluated in retarded time. \\
	The electrons and positrons emit radiation continuously along their track. To get a consistent 
	description	of radio emission in air showers, however, not only contributions due to the deflection of the particles in the magnetic field
	have to be taken into account, but also contributions due to the variation of the number of charged particles. \\
	REAS2 treats radiation processes only along the trajectories, but not at the end or the beginning of trajectories.
	Strictly speaking, this is equivalent to the situation that the particle arrives with the velocity $v\approx c$ given by 
	CORSIKA, enters the Earth's magnetic field where it is deflected 
	and describes a short curved track and finally flies out of the influence of the magnetic field still with a velocity $v\approx c$. 
	\begin{figure}[h!]
		\begin{minipage}[b]{0.47\textwidth}
		\centering
		\includegraphics[width = 0.6\textwidth]{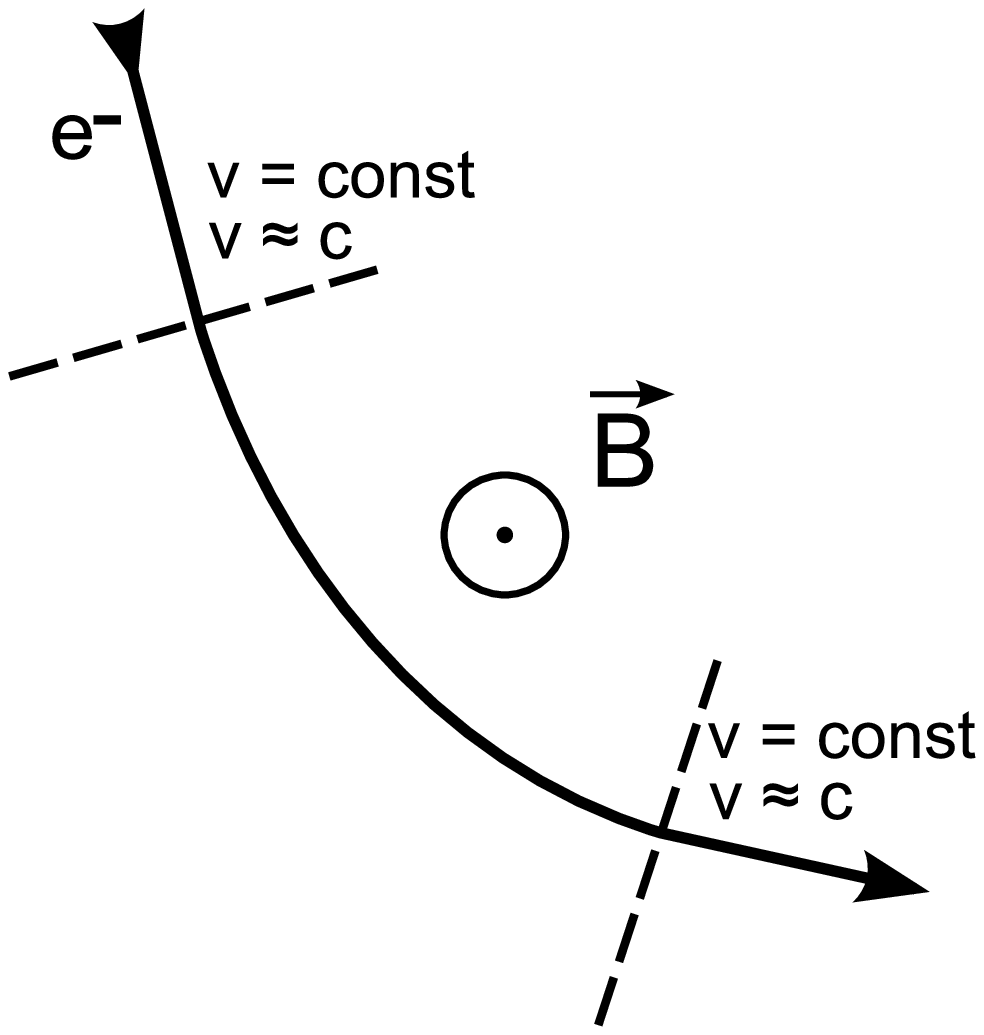}		
		\caption{Sketch of the the trajectory how it was implemented in REAS2.} \label{fig:continuous}
		\end{minipage}
		\hspace{0.6cm}
		\begin{minipage}[b]{0.47\textwidth}
		\centering
		\includegraphics[width = 0.6\textwidth]{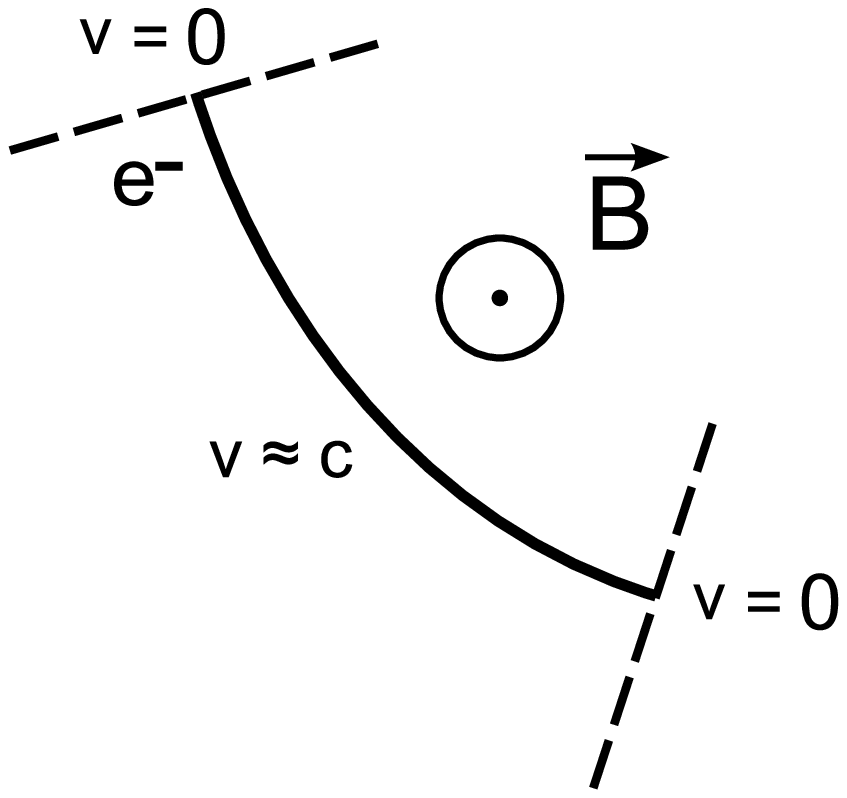}
		\caption{Sketch of the the trajectory with start- and end-point as needed for a consistent description of radio emission from EAS.} \label{fig:cont_short}
		\end{minipage}
	\end{figure}	
	Figure 	\ref{fig:continuous} shows a sketch for such a particle trajectory. It is obvious that this is not describing the real
	situation in an air shower. 
	Consequently, to complement the description in the Monte Carlo code, radiation at the beginning and 
	at the end of each particle trajectory has to be calculated. 
	If at a given atmospheric depth more particle trajectories start
	than end, i.e. the number of charged particles grows, this results in a net contribution. The same is true if the number of 
	charged particles declines, i.e. more particle tracks end than start. Note that a net contribution occurs as well due to the 
	change of the geometrical distribution caused by the spatial separation of the charged particles. \\
	In REAS3, a particle is treated
	as if it was created at rest and became ``instantaneously'' accelerated to $v \approx c$, flew on a short curved track through the
	Earth's magnetic field and got decelerated to rest again (see figure \ref{fig:cont_short}). The acceleration process at the 
	injection as well as the removal of an electron or positron takes place on time scales short in comparison with 
	the frequencies of interest ($\nu_{\mbox{\tiny{observed}}} \leq 100-1000\,$MHz), i.e. $\delta t \ll \frac{1}{\nu_{\mbox{\tiny{observed}}}}$. 
	Hence, only the time-averaged process is of interest, which will give a discrete contribution in contrast to the	
	continuous emission along the curved particle trajectory. To calculate 
	contributions for the start-point, we consider the far field term of the general radiation equation and integrate over the 
	injection time $\delta t$. The ``static'' term of the radiation formula (\ref{eq:radEq}), the velocity field, can be neglected,
	because the ``radiation'' term completely dominates the signal for distances $R$ relevant in practical applications. The 
	relation for the retarded time used for transforming the integral from d$t$ to d$t^{\prime}$ is
	derived from $t^{\prime} = t - R(t^{\prime})/c$. With d$t = (1 -\vec{\beta}\cdot\vec{n})\mathrm{d}t^{\prime} $ the integral is 
	solved as shown in the following calculation: 
	\begin{align}
		\int \vec{E}(\vec{x},t)\md t & =  
		\frac{e}{c} \int_{t_0}^{t_1}\left\vert \frac{ \vec{n}\times [(\vec{n}-\vec{\beta})\times \dot{\vec{\beta}}]} 
		{(1-\vec{\beta}\cdot\vec{n})^3 R}\right\vert_{ret} \md t = \frac{e}{cR} \int_{t'_0}^{t'_1} \frac{ \vec{n}\times 
		[(\vec{n}-\vec{\beta})\times \dot{\vec{\beta}}]} {(1-\vec{\beta}\cdot\vec{n})^2} \md t'\nonumber \\
		& = \frac{e}{cR} \int_{t'_0}^{t'_1} \frac{\md}{\md t}\left(\frac{\vec{n}\times(\vec{n}\times\vec{\beta})} 
		{(1-\vec{\beta}\cdot\vec{n})} \right)\md t' \nonumber\\
		& = \frac{e}{cR} \left[ 1\cdot \frac{\vec{n}\times(\vec{n}\times\vec{\beta})} 
		{(1-\vec{\beta}\cdot\vec{n})}\right]^{t'_1}_{t'_0} -  \frac{e}{cR} \int_{t'_0}^{t'_1} 0 \md t' \nonumber \\
		& = \frac{e}{cR}\left( \frac{\vec{n}\times(\vec{n}\times\vec{\beta})}{(1-\vec{\beta}\cdot\vec{n})}\right) 
		\label{eq:calcStart}
	\end{align}
	Likewise, one gains the electric field for the end-point of the trajectory:	
	\begin{align}
	\int \vec{E}(\vec{x},t)dt = - \frac{e}{cR}\left( \frac{\vec{n}\times(\vec{n}\times\vec{\beta})}{(1-\vec{\beta}\cdot\vec{n})}\right)
	\label{eq:calcEnd}
	\end{align}
	It is important to first transform the integration time into retarded time before calculating the integral. 
	The electric
	fields of the end-points\footnote{In the following we use the term ``end-points'' as a general term for both, start-points and
	end-points, as both are treated in the exact same way.} and of the radiation along the track then have to be added to get a self-consistent
	implementation for modelling radio emission in an air showers.
	
\section{Continuous vs. discrete calculation and incorporation of endpoint contributions}\label{ch:calculations}

\begin{figure}[h!]
		\centering
		\includegraphics[angle = 270 , scale = 0.55]{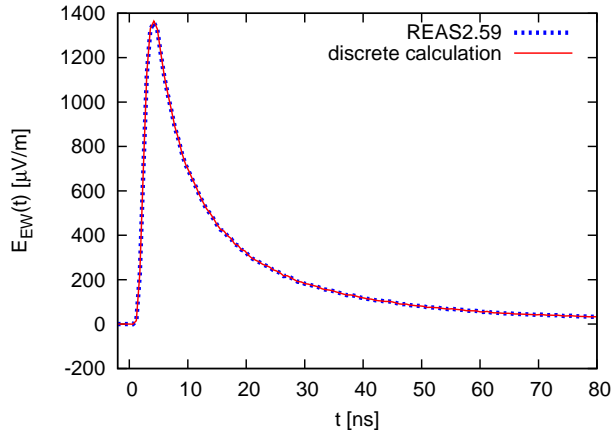}
		\caption{Comparison of discrete and continuous description of radio emission (without endpoint contributions) for an 
		observer 100\,m north of shower core of a vertical $10^{17}$\,eV air shower. The results are identical.} \label{fig:discvscont}
	\end{figure}

	Adding the discrete endpoint contributions to the continuous contributions along the tracks may produce problems.
	The radiation associated with the end-points not only contains the emission due to the tangential acceleration. Due to the change in the
	direction of particle movement between the beginning and the end of the trajectory, radiation associated with the perpendicular acceleration,
	which was so far treated in the continuous description, is also contained. Combining the two descriptions therefore exhibits a risk of double-
	counting. In order to avoid such problems, it is preferable to change the 
	calculation for the continuous contributions along the curved particle tracks to a discrete representation.
	Hence, the chosen representation in REAS3 is completly discrete to ensure that the calculations are self-consistent.
    \begin{figure}[h!]
	\begin{minipage}[h]{0.9\linewidth}
	\centering
    \includegraphics[width=0.25\textwidth]{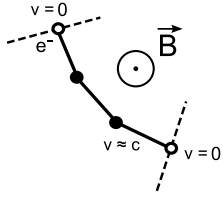}
    \caption{Sketch of the trajectory with a discrete description and end-points.}
   \label{fig:discrete}
	\end{minipage}
	\end{figure}	
	To describe radio emission contributions along the particle trajectories in a discrete picture, the trajectories of the particles are split in
	straight track fragments joined by ``kinks'' in which the velocity of the particles is changing instantaneously. A sketch of this description 
	is given in figure \ref{fig:discrete}. The instantaneous change of velocity at the kinks leads to radiation. With particle
	velocity $\vec{\beta_1}\cdot c$ before and $\vec{\beta_2}\cdot c$ after the kink, radiation for one kink
	of the trajectory is:	 \begin{align}
	 \displaystyle\int\vec{E}(\vec{x},t)\md t &= \int_{t_1}^{t_2}\frac{e}{c} \left\vert \frac{ \vec{n}\times [(\vec{n}-\vec{\beta})\times
	 \dot{\vec{\beta}}]} {(1-\vec{\beta}\cdot\vec{n})^3 R}\right\vert_{ret} \md t = \vec{F}(t_2) - \vec{F}(t_1) \nonumber \\
	 &= \frac{e}{cR}\left(\frac{\vec{n}\times (\vec{n}\times \vec{\beta_2})} {(1-\vec{\beta_2}\vec{n})} \right) - 
	 \frac{e}{cR}\left(\frac{\vec{n}\times (\vec{n}\times \vec{\beta_1})} {(1-\vec{\beta_1}\vec{n})} \right) \label{eq:discrete}
	 \end{align}	 
	 To verify that the continuous and the discrete calculations of emission contributions along the trajectories are equivalent
	 and that both descriptions produce the same results, the REAS2 code was changed to calculate 
	 emission using the discrete approach of straight track segments connected with kinks (without the additional contributions of
	 end-points). Analytically, the equivalence of the two approaches for the frequency domain has been shown by Konstantinov et al. \cite{Konstantinov}. 
	 Several tests with the REAS code have
	 confirmed that the implementation of the discrete description is equivalent to the continuous one as can be seen in figure 
	 \ref{fig:discvscont}. The figure illustrates the equivalence for a vertical proton-induced air shower with primary energy of 
	 $E_p=10^{17}$\,eV. The observer position is 100\,m north of the 
	 shower core. 
	The advantage of the discrete calculation is the consistency of the description of emission contributions along the tracks and 
	emission at the 
	endpoints which is making the incorporation of radiation at the endpoints canonical. 
	To complement the former implementation in the 
	Monte Carlo code with the emission due to the variation of the number of charged particles, it is therefore convenient to use the discrete 
	description. In the discrete picture, contributions at the beginning and the end of a track are just kinks with $\vec{\beta_1} = 0$ and 
	$\vec{\beta_2} = 0$, respectively. This self-consistent emission model has been incorporated in REAS3, taking the radiation at the
	beginning and the end of a trajectory 
	as well as along the curved trajectory into account. The obtained results for the radio signal is discussed in 
	section \ref{ch:results}, but first the numerical stability will be demonstrated.

\section{Numerical stability}\label{ch:crosschecks}

\begin{figure}[h!]
		\begin{minipage}[b]{0.45\linewidth}
		\centering
		\begin{overpic}[angle = 270 , scale = 0.45]{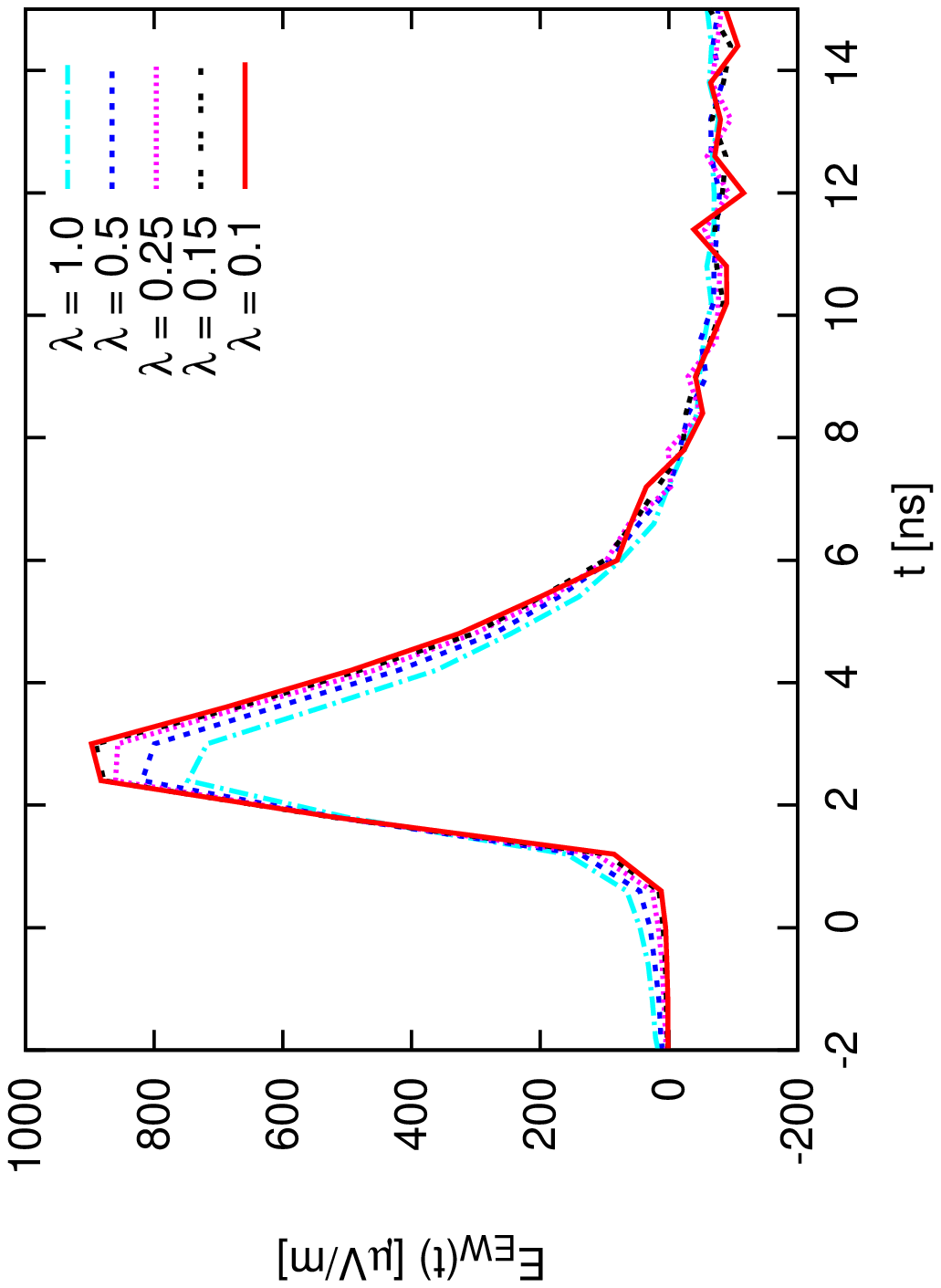}
		\put(45,60){\scriptsize{100\,m\,-\,N}}
		\end{overpic}		
		\end{minipage}
		\hspace{0.6cm}
		\begin{minipage}[b]{0.45\linewidth}
		\centering
		\begin{overpic}[angle = 270 , scale = 0.45]{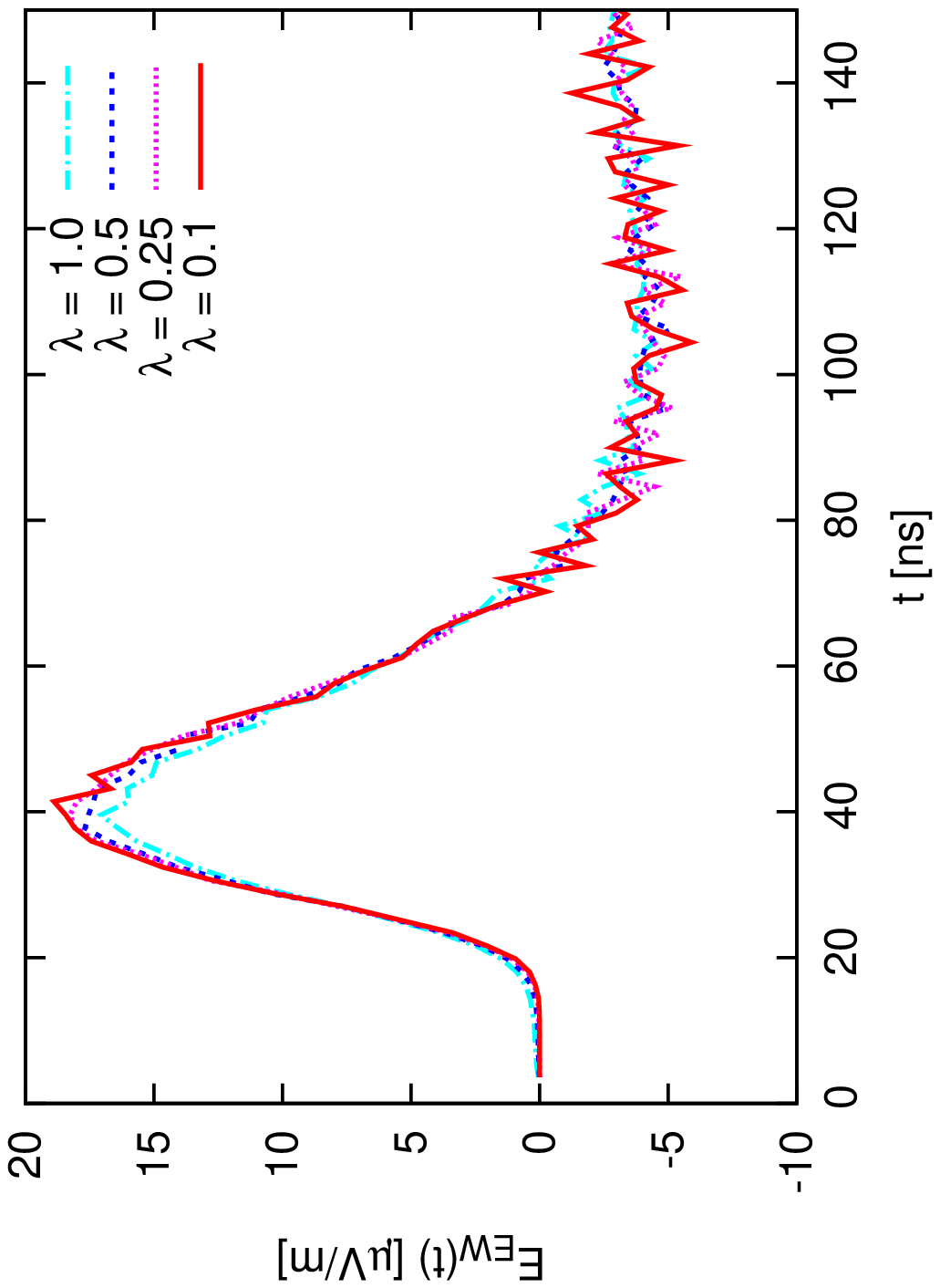}
		\put(45,60){\scriptsize{400\,m\,-\,N}}
		\end{overpic}				
		\end{minipage}
		\caption{Influence of different values of the technical parameter $\lambda$. Left: observer 100\,m north of shower core. Right:
		observer 400\,m north of shower core.}\label{fig:lambda}
	\end{figure}

	As already mentioned in chapter \ref{ch:algorithm}, particle trajectories are represented by ensembles of several shorter, 
	unrelated trajectories in the code. This ensures that the phase-space distribution of the analytically propagated particles stays consistent
	with the underlying	particle distribution without having to treat energy losses during the propagation explicitly (cf. section 4 of
	\cite{HuegeUlrichEngel2007a}). The length of the short segments is controlled by a parameter $\lambda$.
	This parameter $\lambda$ determines the length of the short tracks into which the real trajectories of the 
	particles are divided (it does not denote the sampling density of kinks on a track).
	If $\lambda$ is chosen inadequately large, the discrepancies between the particle distributions recreated in REAS and the 
	distributions histogrammed in CORSIKA are getting too large. Thus, to avoid these discrepancies $\lambda$ has to be chosen small enough.	
	In REAS2, it was recommended to set this parameter to $\unit[1]{\frac{g}{cm^2}}$, even though it is just
	a technical parameter, i.e. the result does not depend on the exact value of $\lambda$ as long as $\lambda$ is set small enough. 
	Also, in REAS3 the tracklength of the single short trajectories should have no influence on the physics results 
	as long as it is chosen small enough. This cross check was made for a vertical proton-induced air shower with $E_p=10^{17}$\,eV. 
	As shown in figure \ref{fig:lambda} for
	an observer 100\,m north of the shower core, the result converges with decreasing values of $\lambda$. The same result is obtained
	for an observer 400\,m north of the shower core which is displayed in the right column of figure \ref{fig:lambda}. The comparison of both
	observer positions demonstrates that far away from the shower core, the result converges faster than close to the shower core. 
	For $\lambda=\unit[0.1]{\frac{g}{cm^2}}$, a stable result is obtained in both cases. 
	For observer positions close to the shower core as for 100\,m distance, it is also visible that for larger values of
	$\lambda$, e.g. for	$\lambda=\unit[1.0]{\frac{g}{cm^2}}$, there is already a finite radio contribution for negative	times. These
	unphysical contributions appear if partciles are created with symmetric trajectories around the place of particle generation and $\lambda$ is
	chosen inadequately large. For small enough values of $\lambda$, however, the symmetric trajectories ensure a faster convergence of the result
	than the asymmetric trajectories used in REAS2.\\
	With smaller parameters of $\lambda$, however, the high frequency noise increases as well as the computing time.
	This first effect is larger for observers far away as can be seen in figure \ref{fig:lambda}.
	To optimize the calculation in REAS3, the path depth $\lambda$ of
	the electron and positron trajectories is therefore chosen as a dynamical parameter depending on the lateral distance of the observer. 
	Therefore, it is recommended to set $ \lambda $ to $\unit[0.1]{\frac{g}{cm^2}} $ at the core and increase this value
	linearly every 100\,m as it is done by default in REAS3.
	The advantage of this implementation is on the one hand to gain stable results for all observers and on the other hand to 
	avoid high frequency noise for observers at larger lateral distances. 
	The difference in the recommended value of $\lambda$ in REAS2 and REAS3 is due to the fact that the emission model in 
	REAS3 requires a precise description of the particle momenta during the propagation, which requires a more fine-grained
	treatment.

\section{Results}\label{ch:results}
	
	In chapter \ref{ch:crosschecks}, we have verified that REAS3 is producing stable results and that the endpoints 
	were implemented correctly. 
	In chapter \ref{REAS2vsREAS3}, we focus on the results obtained with REAS3 and in section
	\ref{chargeexcess} we discuss the influence of the charge excess in air showers on the radio emission. 
	In both chapters, the radio emission was 
	calculated for a proton-induced vertical air shower with primary energy of $E_p=10^{17}$eV. The shower 
	itself was generated with CORSIKA 6.7 and COAST. The positions of the observers were chosen at sea level with different lateral distances 
	and relative observer orientations to the shower core. For the simulations with REAS2 and REAS3, identical showers were taken.
	This is easily possible because the histogramming approach allows an easy seperation between the air shower modelling and the 
	radio emission calculation.
	
	\subsection{REAS2 vs. REAS3}\label{REAS2vsREAS3}
	
\begin{figure}[h!]	
		\begin{minipage}[b]{0.45\linewidth}
		\centering
		\begin{overpic}[angle = 270 , scale = 0.45]{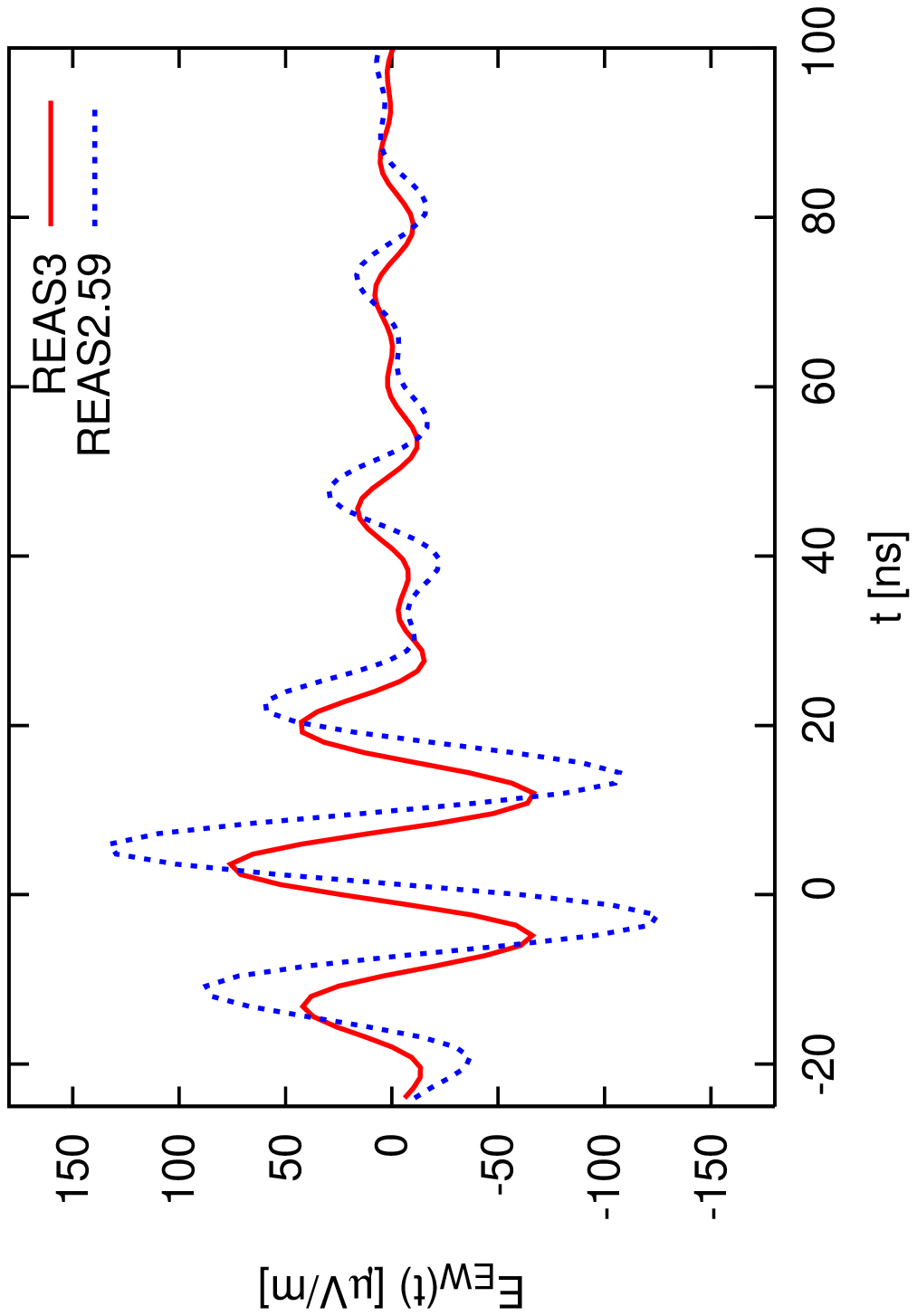}		
		\put(35,60){\scriptsize{100\,m\,-\,N}}
		\end{overpic}
		\end{minipage}
		\hspace{0.8cm}
		\begin{minipage}[b]{0.45\linewidth}
		\centering
		\begin{overpic}[angle = 270 , scale = 0.45]{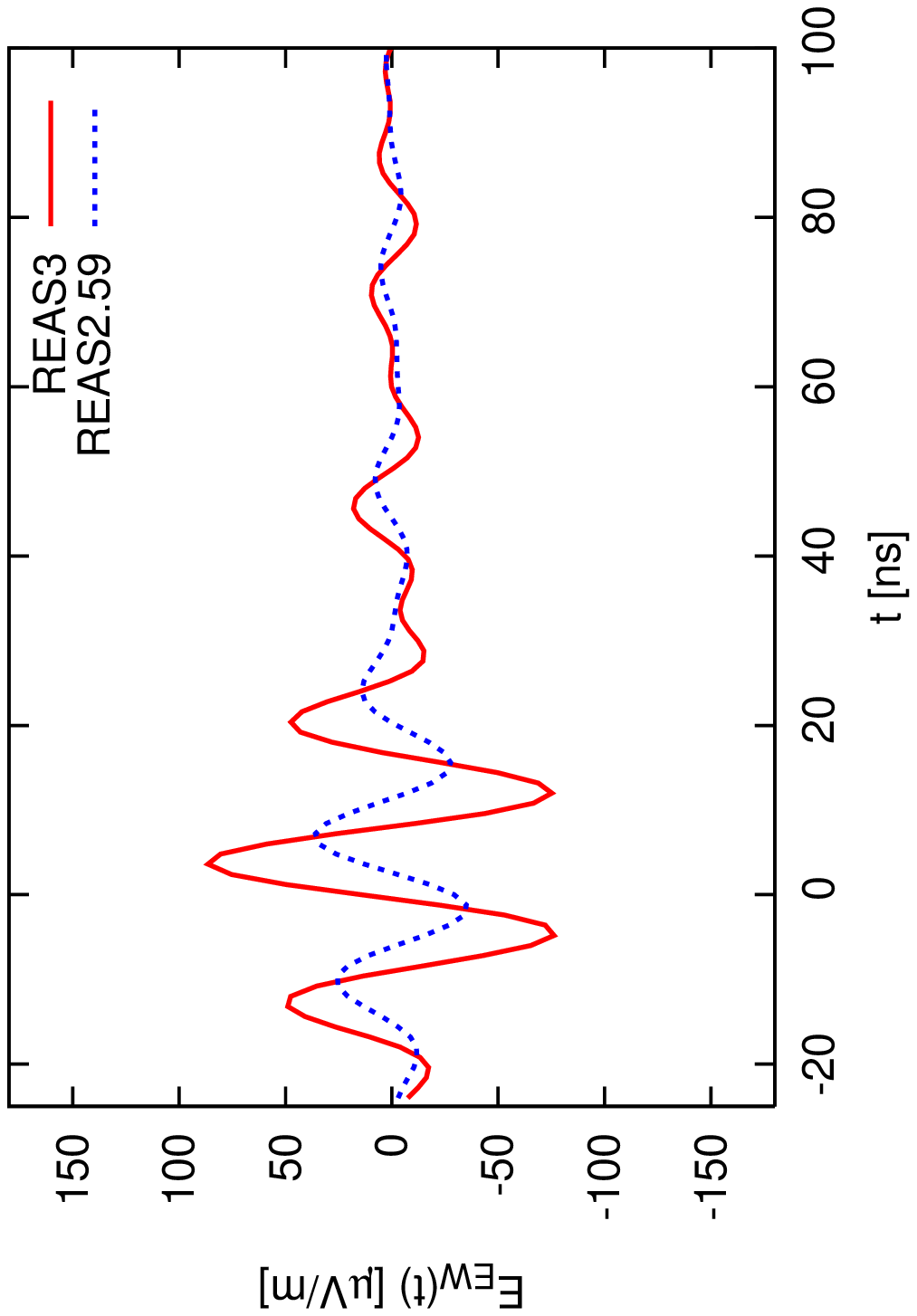}
		\put(35,60){\scriptsize{100\,m\,-\,E}}
		\end{overpic}
		\end{minipage}
		\caption{Filtered pulses (a simple rectangular 43-76\,MHz bandpass filter is used) for observers with 100\,m distance to the shower
		core. Left: the observer is north of the shower core. Right: the observer is east of the shower core.}\label{fig:filtered}
	\end{figure}

\begin{figure}[h!]
		\begin{minipage}[b]{0.27\linewidth}
		\centering
		\includegraphics[angle = 270 , scale = 0.33]{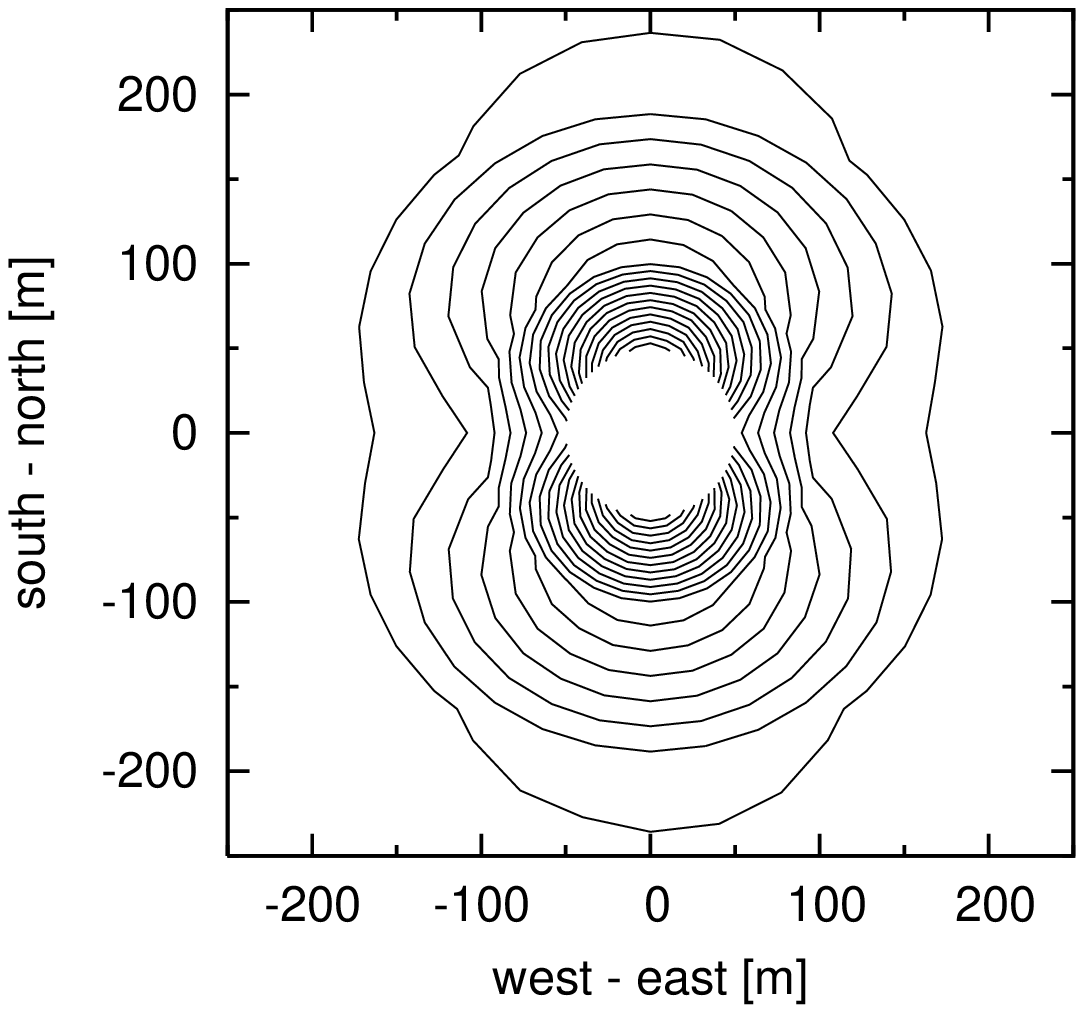}		
		\includegraphics[angle = 270 , scale = 0.33]{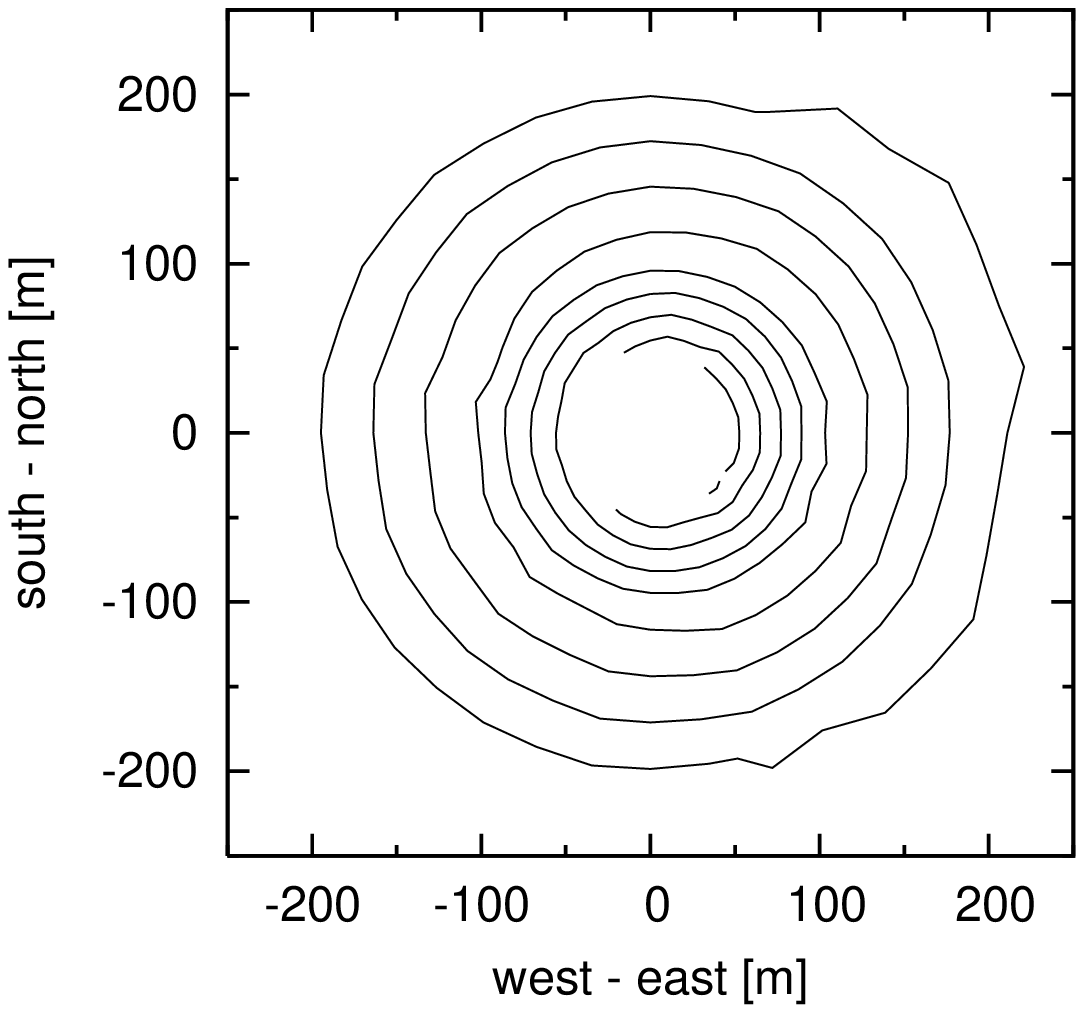}		
		\end{minipage}
		\hspace{0.4cm}
		\begin{minipage}[b]{0.27\linewidth}
		\centering
		\includegraphics[angle = 270 , scale = 0.33]{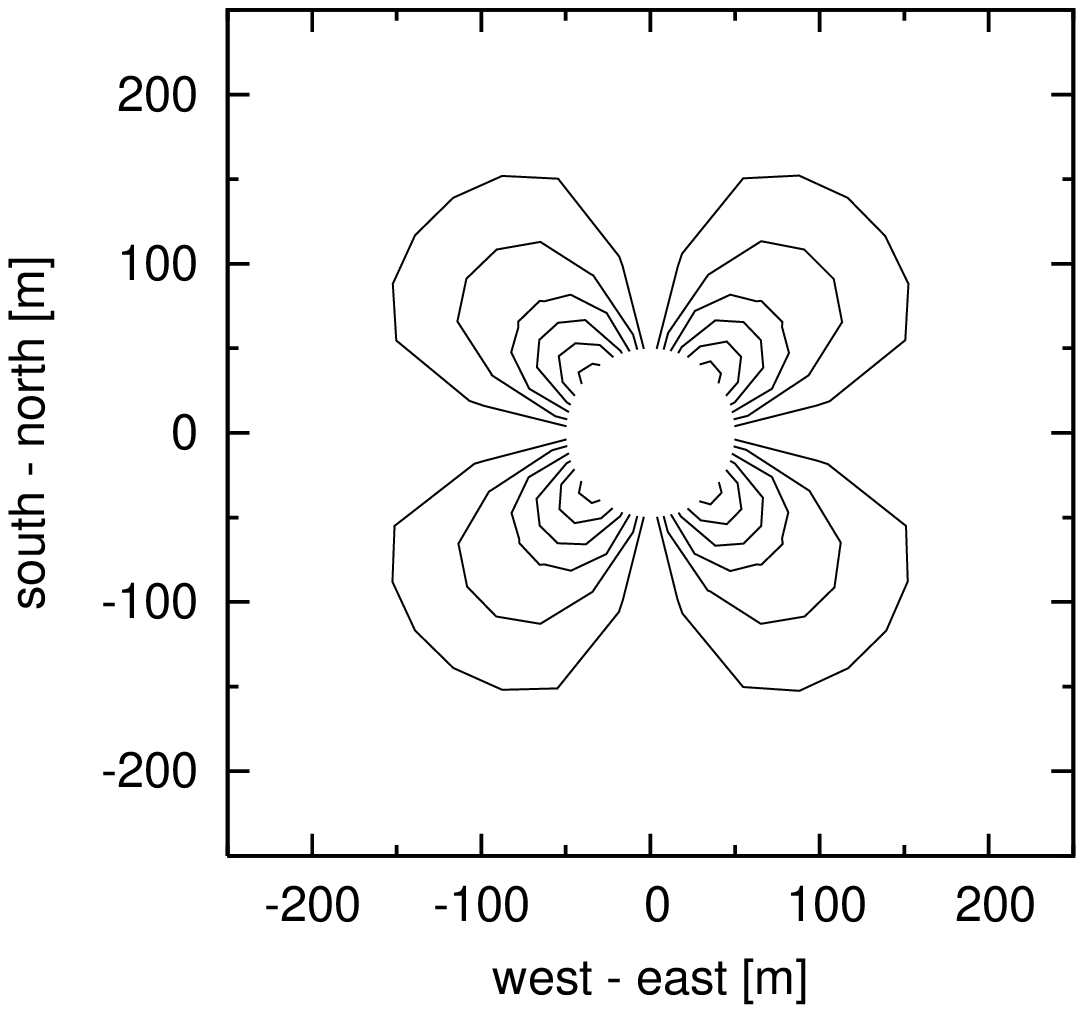}
		\includegraphics[angle = 270 , scale = 0.33]{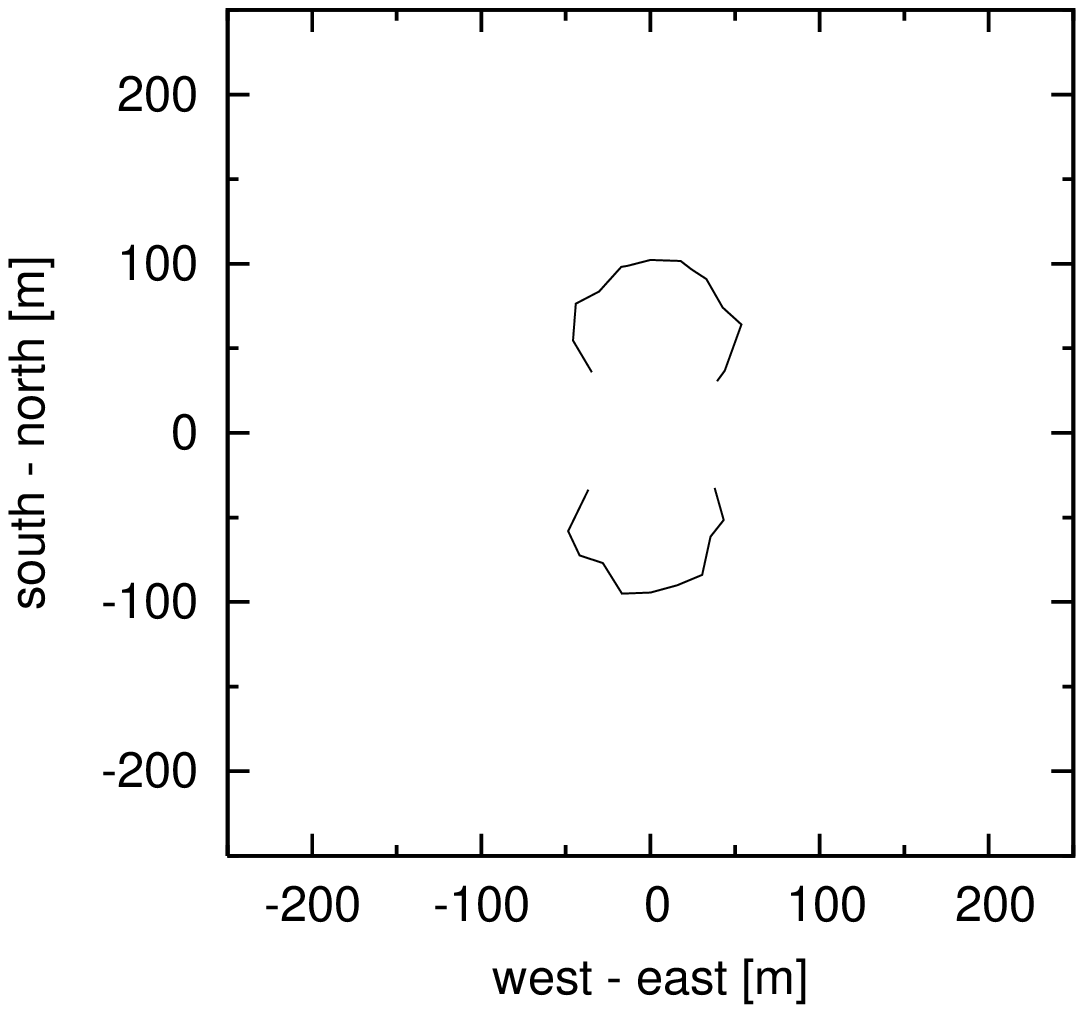}
		\end{minipage}
		\hspace{0.4cm}
		\begin{minipage}[b]{0.27\linewidth}
		\centering
		\includegraphics[angle = 270 , scale = 0.33]{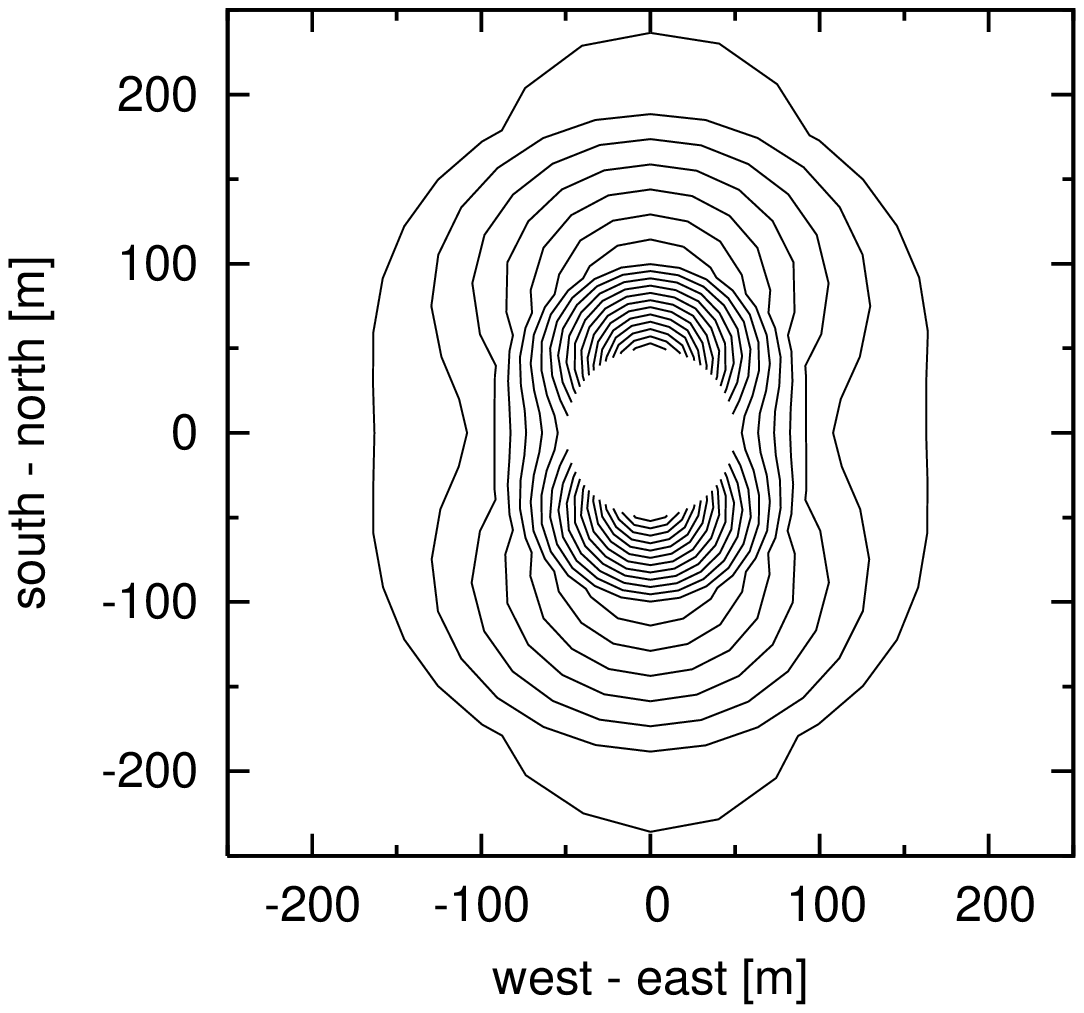}		
		\includegraphics[angle = 270 , scale = 0.33]{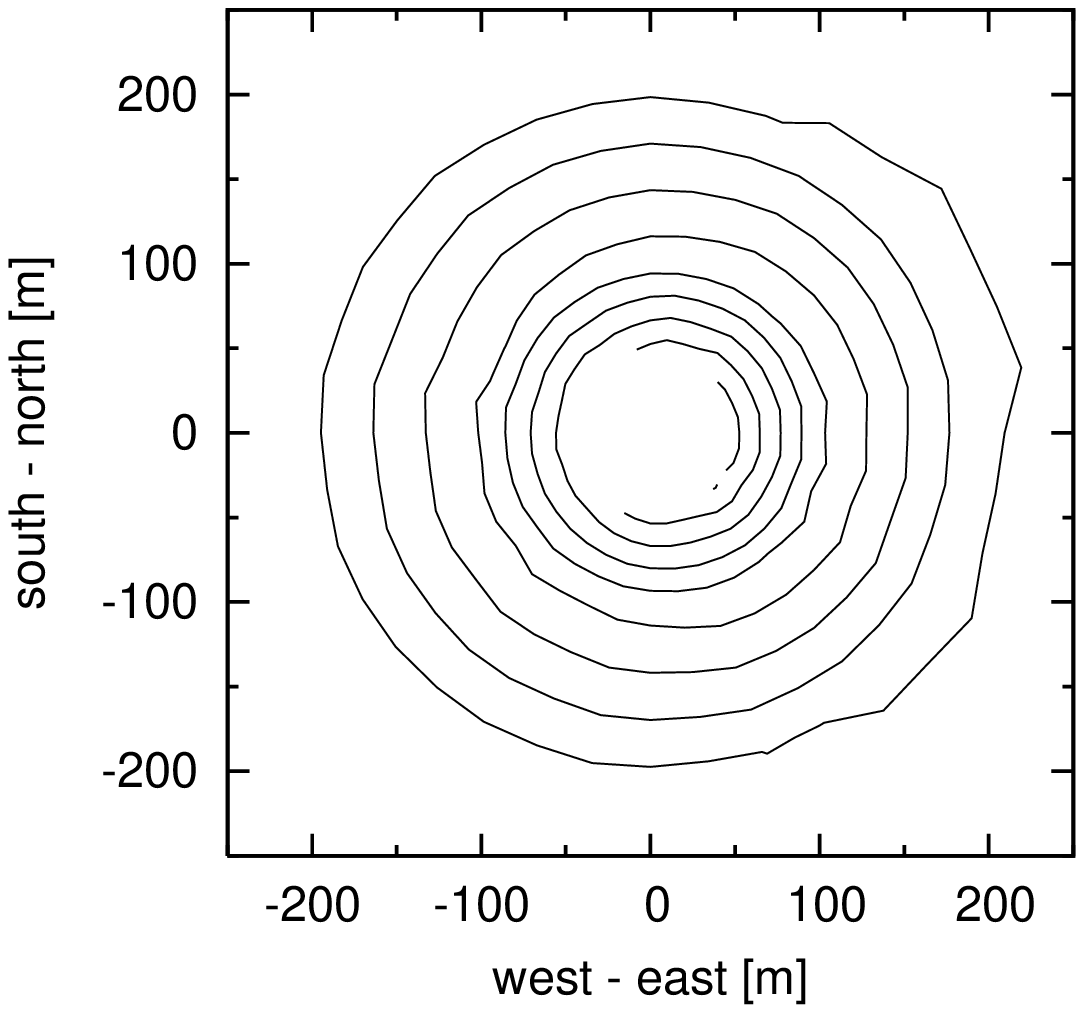}		
		\end{minipage}
		\caption{Contour plots of the 60\,MHz field strength for emission from a $10^{17}$\,eV vertical air shower.
		From left to right:	total field strength, north-south and east-west polarisation component. 
		Contour levels are 0.1\,$\mu$Vm$^{-1}$MHz$^{-1}$ apart. The closest position of the simulated observers to the shower core
		is 50\,m. Upper row: REAS2. Lower row: REAS3} \label{fig:contour}
	\end{figure}

	To study the changes introduced by the implementation of emission due to the variation of the number of charged
	particles, a simple, vertical shower geometry was chosen as specified above. The magnetic field was taken as horizontal with a field 
	strength of 0.23\,Gauss to get a geomagnetic angle of 90$^\circ$.    
	\begin{figure}[h!]	
    	\hspace{-1.3cm}
		\begin{minipage}[b]{0.45\textwidth}
		\centering
		\begin{overpic}[angle = 270 , scale = 0.45]{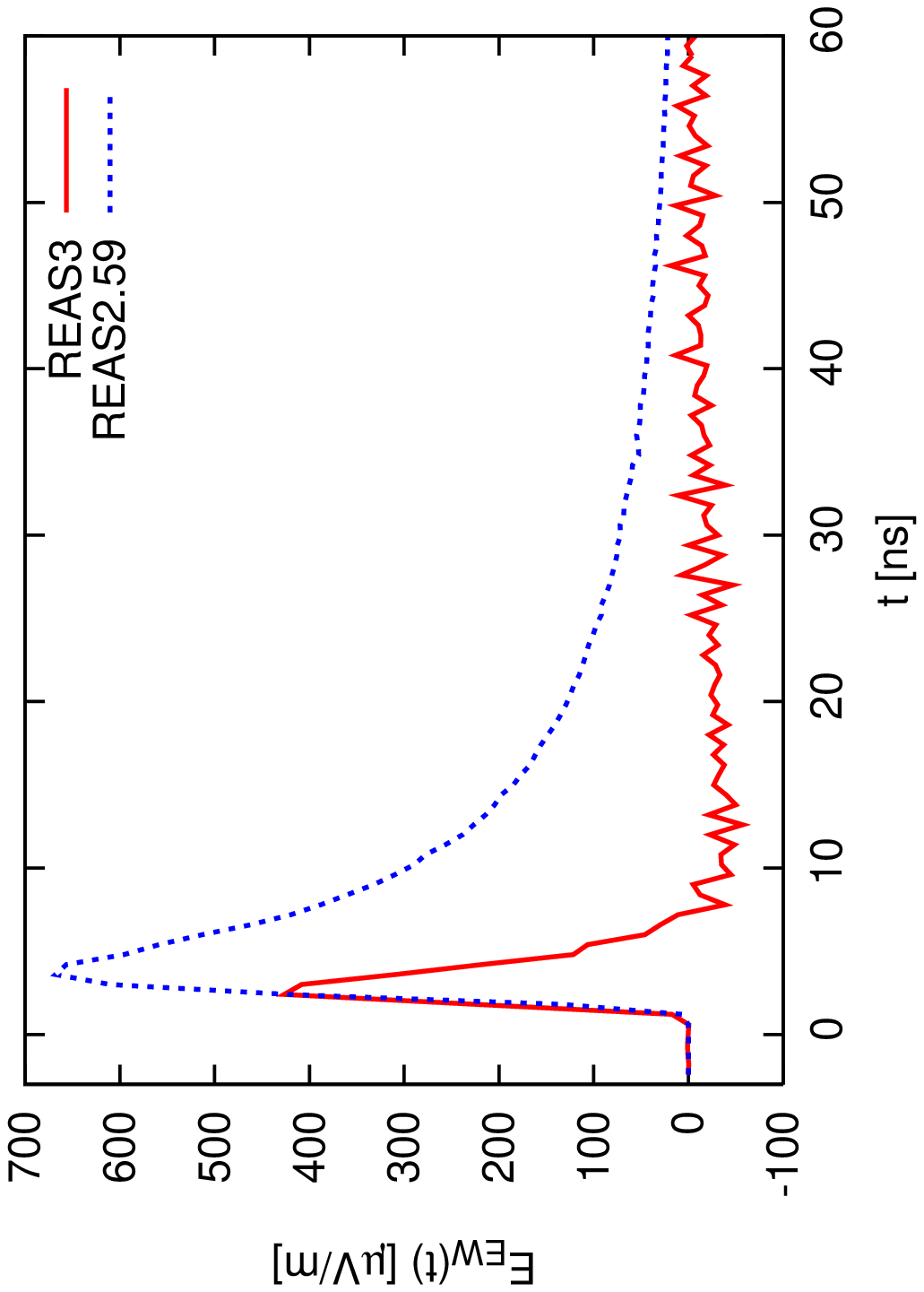}
		\put(35,60){\scriptsize{100\,m\,-\,N}}
		\end{overpic}
		\begin{overpic}[angle = 270 , scale = 0.45]{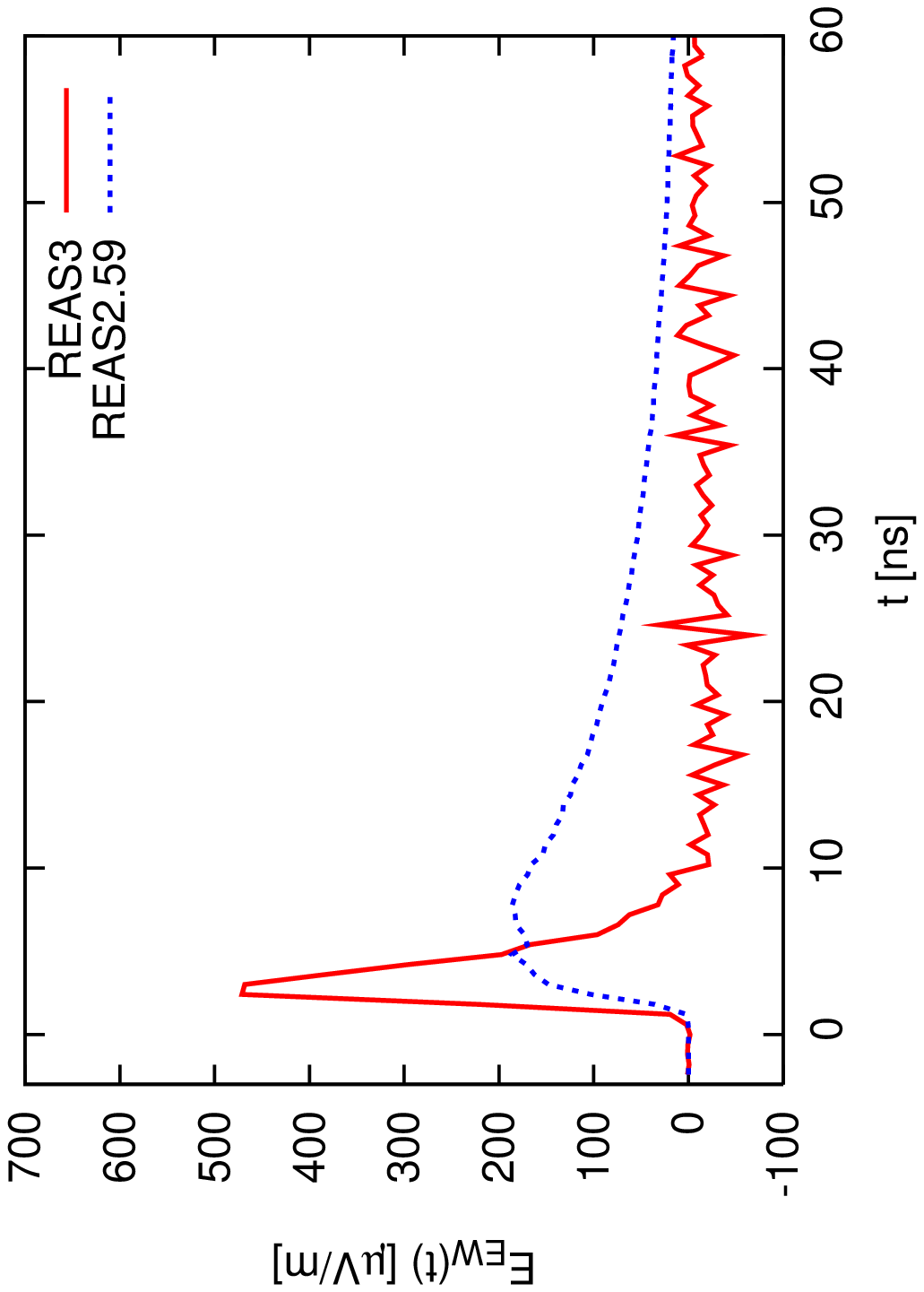}		
		\put(35,60){\scriptsize{100\,m\,-\,E}}
		\end{overpic}
		\begin{overpic}[angle = 270 , scale = 0.45]{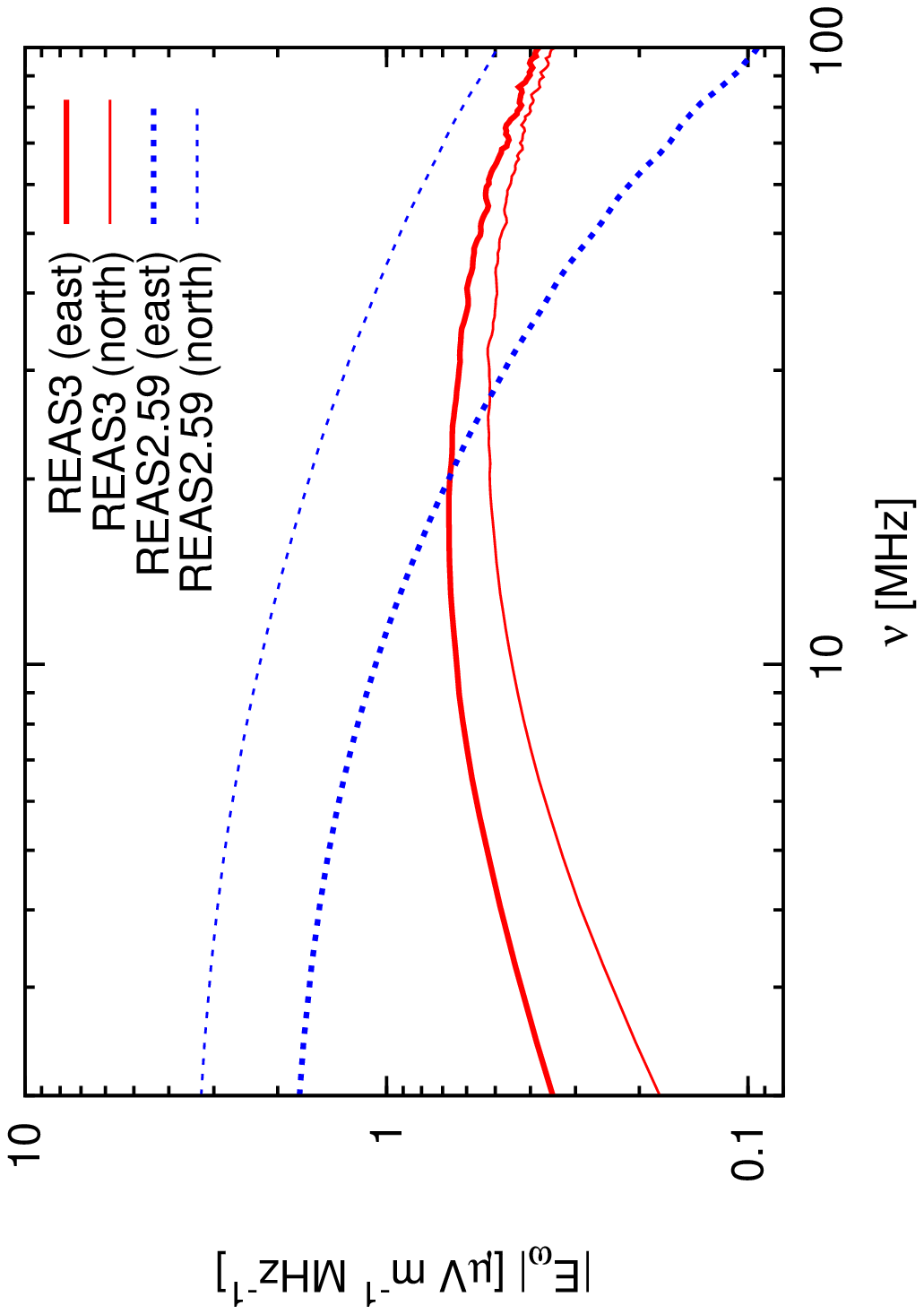}		
		\put(35,60){\scriptsize{100\,m}}
		\end{overpic}
		\end{minipage}
		\hspace{1.0cm}
		\begin{minipage}[b]{0.45\textwidth}
		\centering
		\begin{overpic}[angle = 270 , scale = 0.45]{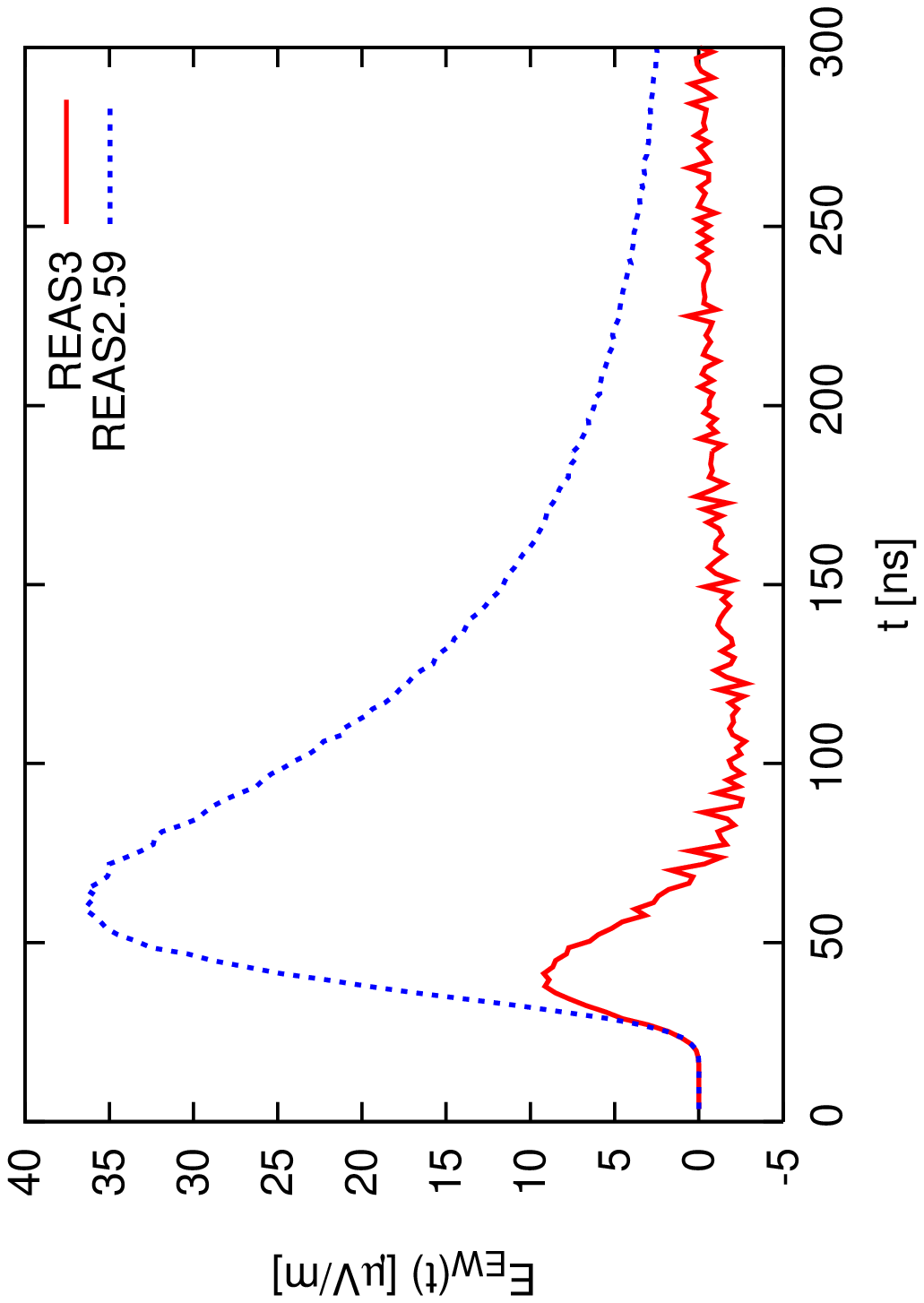}		
		\put(35,60){\scriptsize{400\,m\,-\,N}}
		\end{overpic}
		\begin{overpic}[angle = 270 , scale = 0.45]{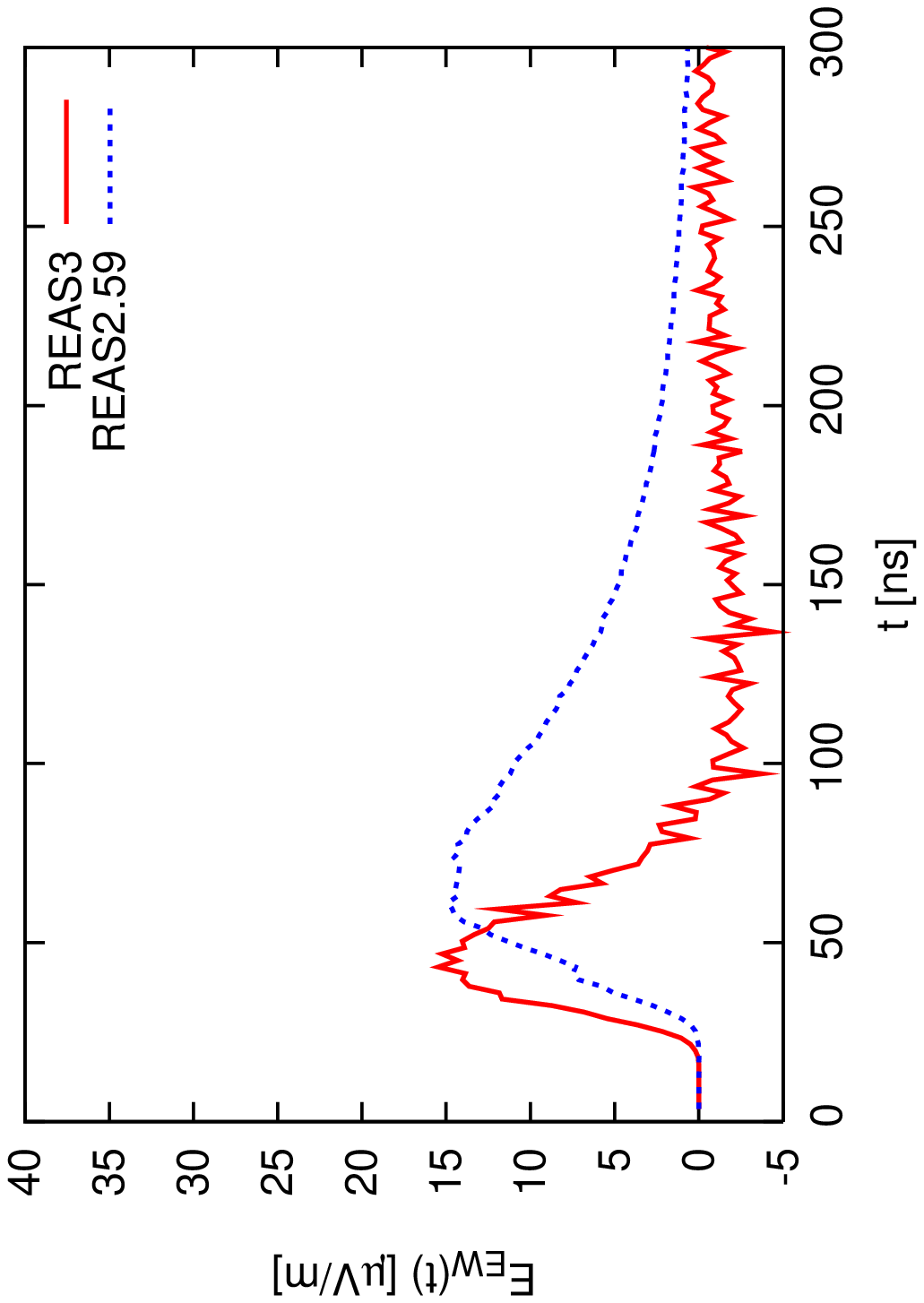}
		\put(35,60){\scriptsize{400\,m\,-\,E}}
		\end{overpic}		
		\begin{overpic}[angle = 270 , scale = 0.45]{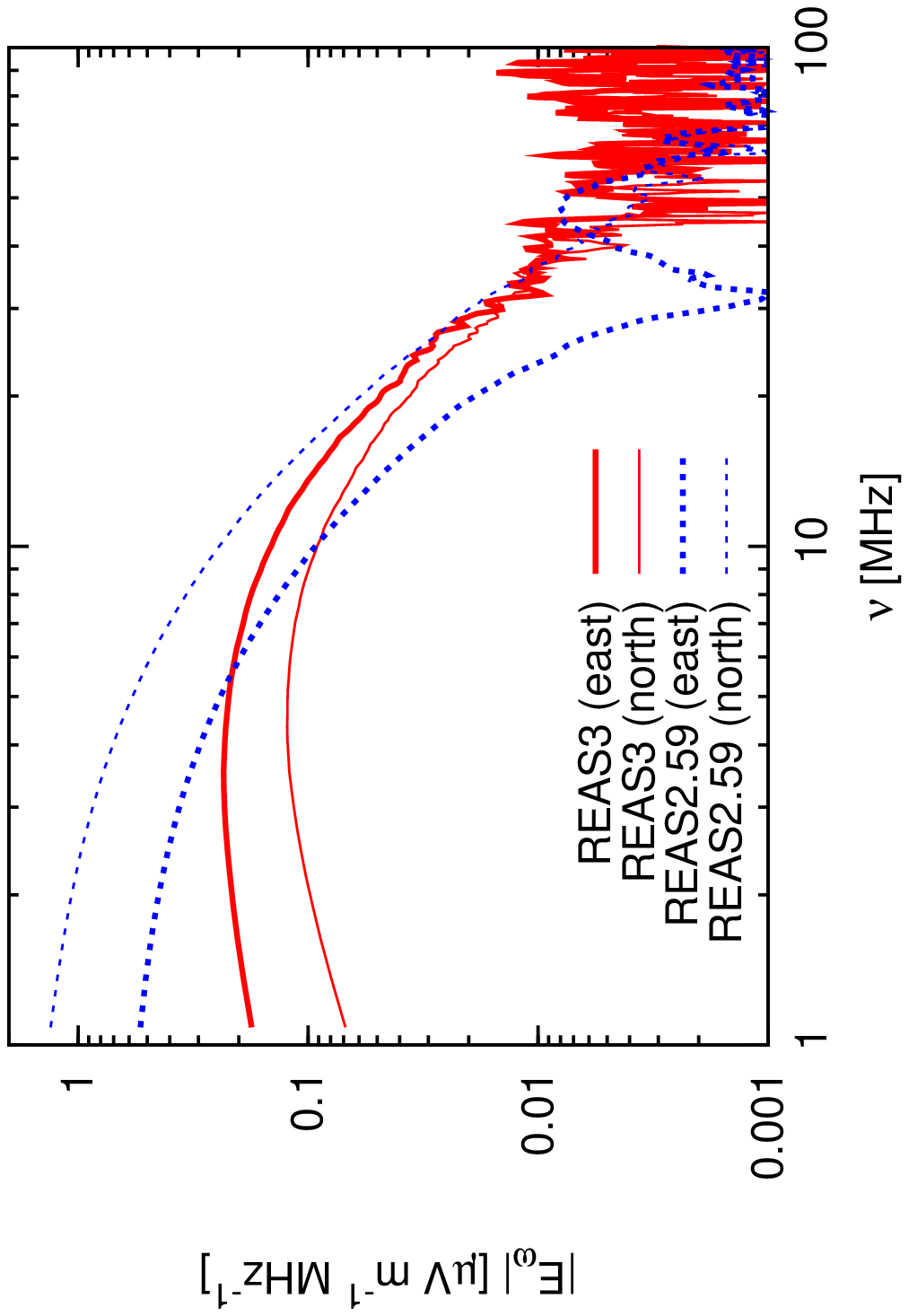}
		\put(45,60){\scriptsize{400\,m}}
		\end{overpic}
		\end{minipage}
		\caption{Upper row: raw pulses for observers 100\,m (left) and 400\,m (right) north of shower core. Middle row: 
		raw pulses for observers 100\,m (left) and 400\,m (right) east of
		shower core. Lowest row: frequency spectra for observers 100\,m (left) and 400\,m (right) east and north of shower core. For the raw pulses
		the east-west polarisation of the electric field is shown, whereas for the frequency spectra the total spectral field is shown.}\label{fig:pulses}
	\end{figure}
	Figure \ref{fig:pulses} shows a comparison between unfiltered (i.e. unlimited bandwidth) pulses and frequency spectra of 
	REAS2 and REAS3 for observers with different positions with respect to the shower core. 
	The spectra show the total field strength for two azimuthal	observer directions: the thick line displays the total field strength for an
	observer east of the shower core and the thin line for an observer north of the shower core.
	It is clear that the strength of the pulses as well as the time structure of the pulses has changed, 
	while the changes in the pulse	amplitudes are dependent on	the observer azimuth angle.
	The time structure of the pulses has changed from unipolar to bipolar. This can also be seen in the frequency spectra (lowest row
	of figure \ref{fig:pulses}) where in case of REAS3, the spectral field strengths drop to zero for frequency zero, as it was the case in the
	analytical implementation \cite{HuegeFalcke2003a}.
	It can be argued from basic physical arguments that the spectral field strength has to drop to zero for small frequencies because
	the source of the coherent emission exists only for a finite time in a finite region of space and thus the the zero-frequency component of the
	emission, which corresponds to an infinite time-scale, can contain no power (cf. \cite{ScholtenWernerArena2008}).\\
	While the emission 
	pattern of REAS2 was azimuthally asymmetric, REAS3 simulations are much more symmetric, as is expected for the given shower geometry. The 
	increased symmetry can also be seen in the frequency spectra 
	where the characteristics of the spectral field strengths with frequency are getting very similar for the two different azimuthal
	observer positions in case of REAS3.
	The remaining asymmetry will be discussed in section \ref{chargeexcess}. To get a more general impression of the changes in the amplitudes of
	the signal with REAS3, the lateral dependence of the emission is studied. 
	\begin{figure}[h!]
		\begin{minipage}[b]{0.45\linewidth}
		\centering
		\begin{overpic}[angle = 270 , scale = 0.45]{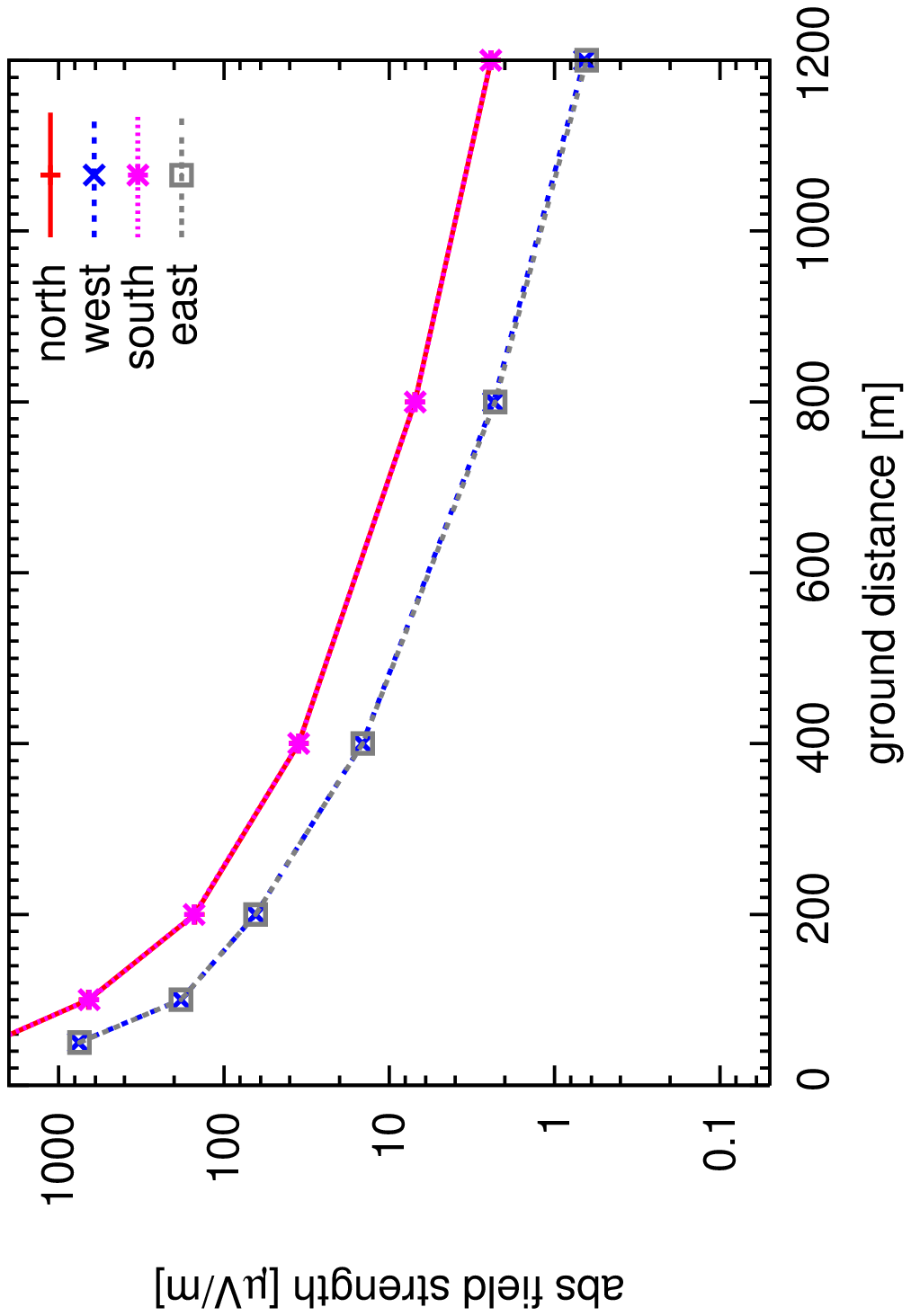}
		\put(35,60){\scriptsize{REAS2}}
		\end{overpic}
		\end{minipage}
		\hspace{0.6cm}
		\begin{minipage}[b]{0.45\linewidth}
		\centering
		\begin{overpic}[angle = 270 , scale = 0.45]{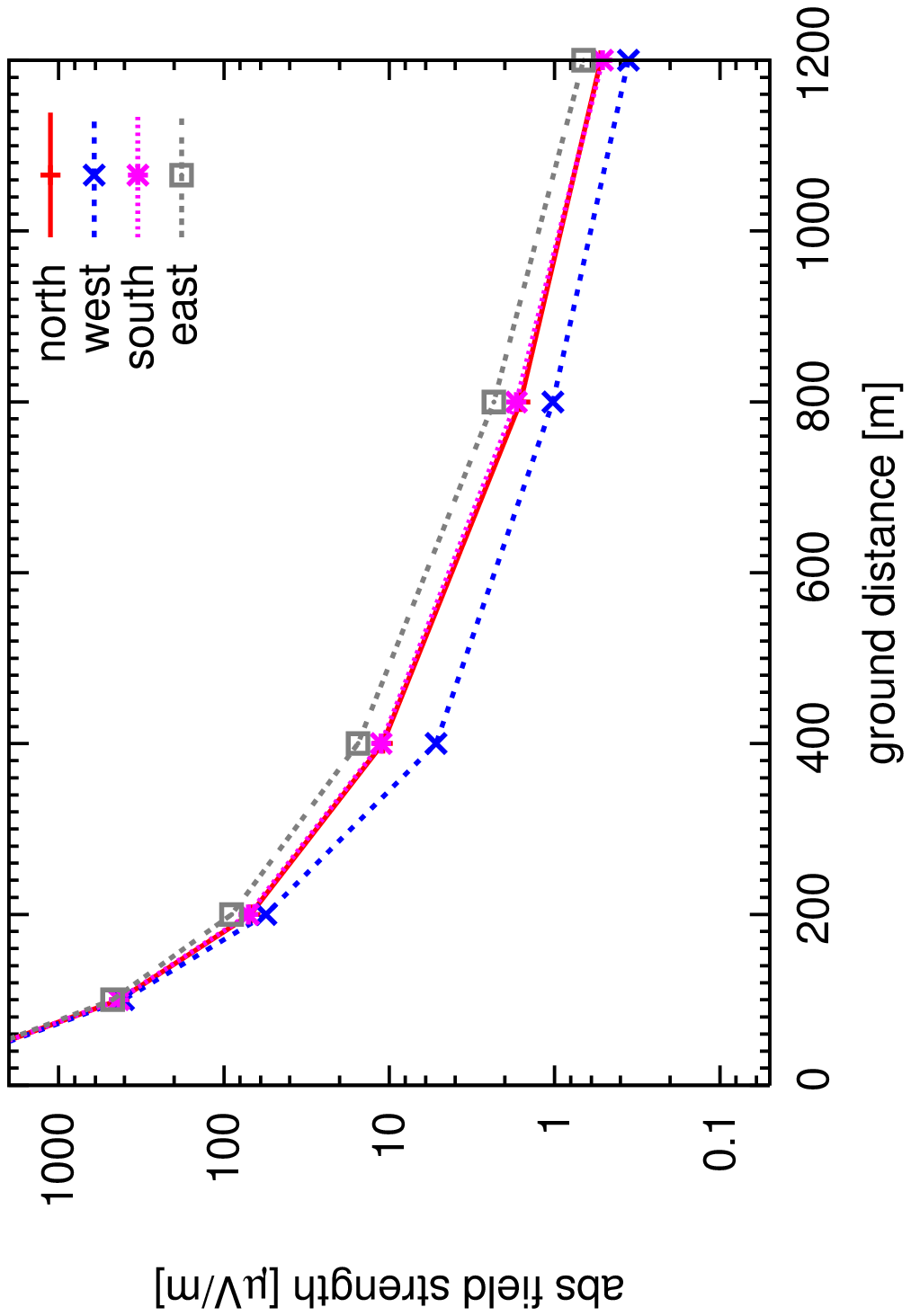}
		\put(35,60){\scriptsize{REAS3}}
		\end{overpic}
		\end{minipage}
		\caption{Lateral dependence with full bandwidth amplitudes. Left: REAS2. Right: REAS3.}
		\label{fig:lateral}
	\end{figure}		
	Figure \ref{fig:lateral} shows this dependence for the unfiltered, full bandwith amplitudes. In REAS2, there was a much stronger
	signal for observers in the north or south than for observers in the east or west. In REAS3, the signal pattern is much more
	symmetric, but obervers in the eastern direction measure a higher absolute field strength and observers in the west see a lower
	field strength than observers with other azimuthal positions. In general, the amplitudes of the
	field strength got lower from REAS2 to REAS3, while the changes for observers north and south of the shower core are
	much larger than for observers in the east or the west. \\ 
	To compare simulations with experimental data (e.g. of LOPES), the REAS simulations have to be filtered to a finite observing
	bandwidth. This is done by the helper application 
	REASPlot which is included in the REAS3 package. 
	In this paper, REASPlot was used with an idealised rectangular 43\,MHz-76\,MHz bandpass filter which can lead to acausal contributions
	at negative times due to the idealisation of the filter.
	In figure \ref{fig:filtered}, the filtered pulses for observers with lateral distance of 100\,m east and north of the 
	shower core are shown.
	The quick oscillations are determined by the selected filter bandwidth.
	The differences in the pulse strengths are much smaller for the filtered than for the raw pulses because the strongest changes between
	REAS2 and REAS3	occur at low frequencies (Please note that figure \ref{fig:pulses} is plotted on a log-log scale). In general, e.g. for different geometries,
	the differences in the amplitudes of the filtered pulses	can be larger than in this example (cf. \cite{LudwigHuegeArena2010}).\\
	Again, the increased azimuthal symmetry of REAS3 is observable. To get an overall impression of the change from 
	REAS2 to REAS3 and the influence of the net charge excess on the east-west asymmetry, we show the contour plots of the 60\,MHz 
	field strength in figure \ref{fig:contour}.
	The ellipticity which is seen in REAS2 is replaced by a nearly azimuthally symmetric pattern in REAS3. 
	The ``clover'' pattern seen earlier in the north-south polarisation component is no longer visible.	In the vertical polarisation 
	component (not shown here), there is no significant flux for either of the two simulations, as expected for a vertical shower. In the contour
	plots of REAS3, the east-west asymmetry is visible as well. The resulting REAS3 emission pattern can be interpreted as a superposition of a 
	circularly symmetric contribution with a $ \vec{v}\times\vec{B} $ (in this case thus pure east-west) polarisation and a radially polarised
	emission contribution caused by the time-varying net charge excess. \\
	In summary, the incorporation of radiation due to the variation of the number of charged particles in the form of endpoint
	contributions results in a clear change from REAS2 to REAS3. 
	The revised, self-consistent implemented model in the Monte Carlo code REAS3
	predicts bipolar pulses with a mostly symmetrical emission footprint. In addition to these changes, a further
	emission contribution arises from the variation of the net charge excess, which explains the remaining asymmetries. This
	contribution will be discussed in the following section.
		
	\subsection{Discussion of charge excess emission}\label{chargeexcess}

\begin{figure}[h!]
		\begin{minipage}[b]{0.27\linewidth}
		\centering
		\includegraphics[angle = 270 , scale = 0.33]{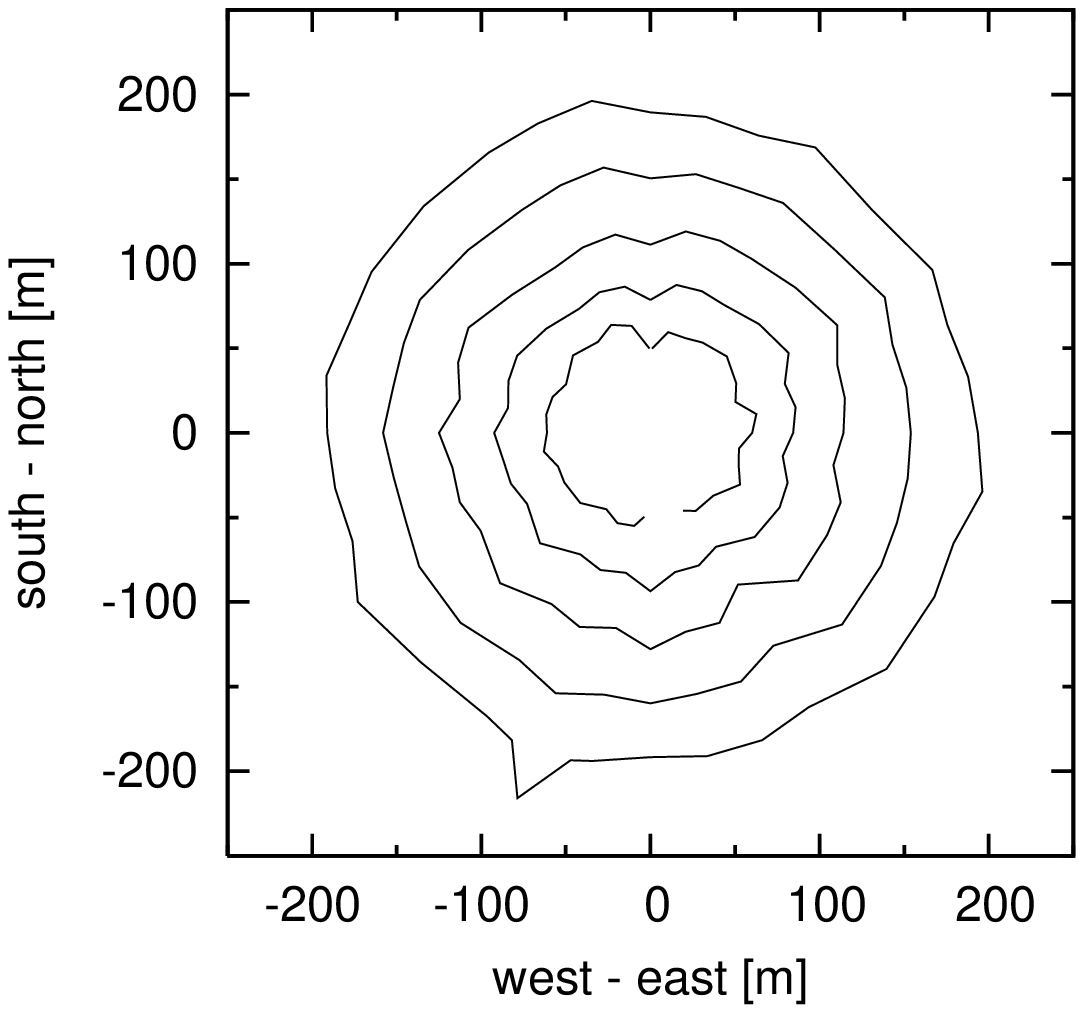}		
		\end{minipage}
		\hspace{0.4cm}
		\begin{minipage}[b]{0.27\linewidth}
		\centering
		\includegraphics[angle = 270 , scale = 0.33]{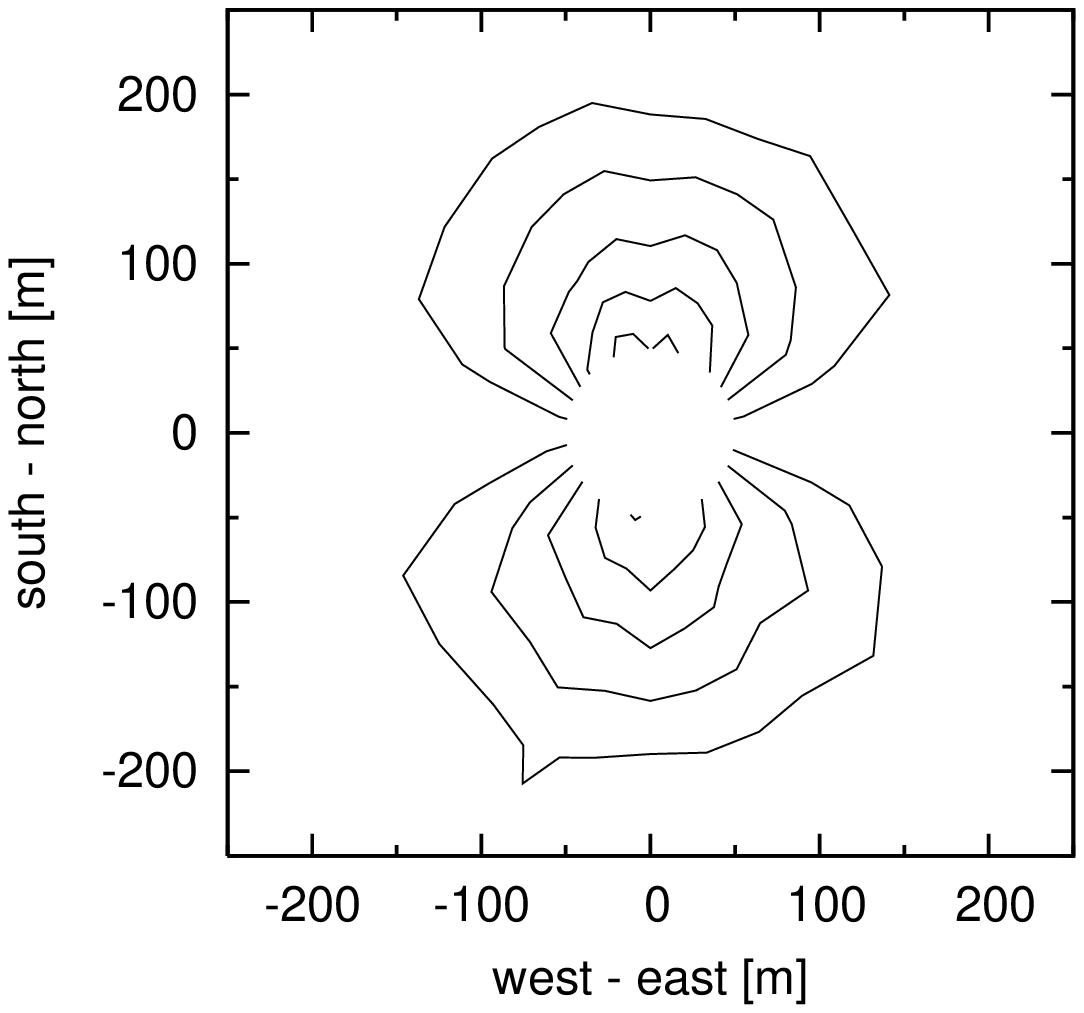}
		\end{minipage}
		\hspace{0.4cm}
		\begin{minipage}[b]{0.27\linewidth}
		\centering
		\includegraphics[angle = 270 , scale = 0.33]{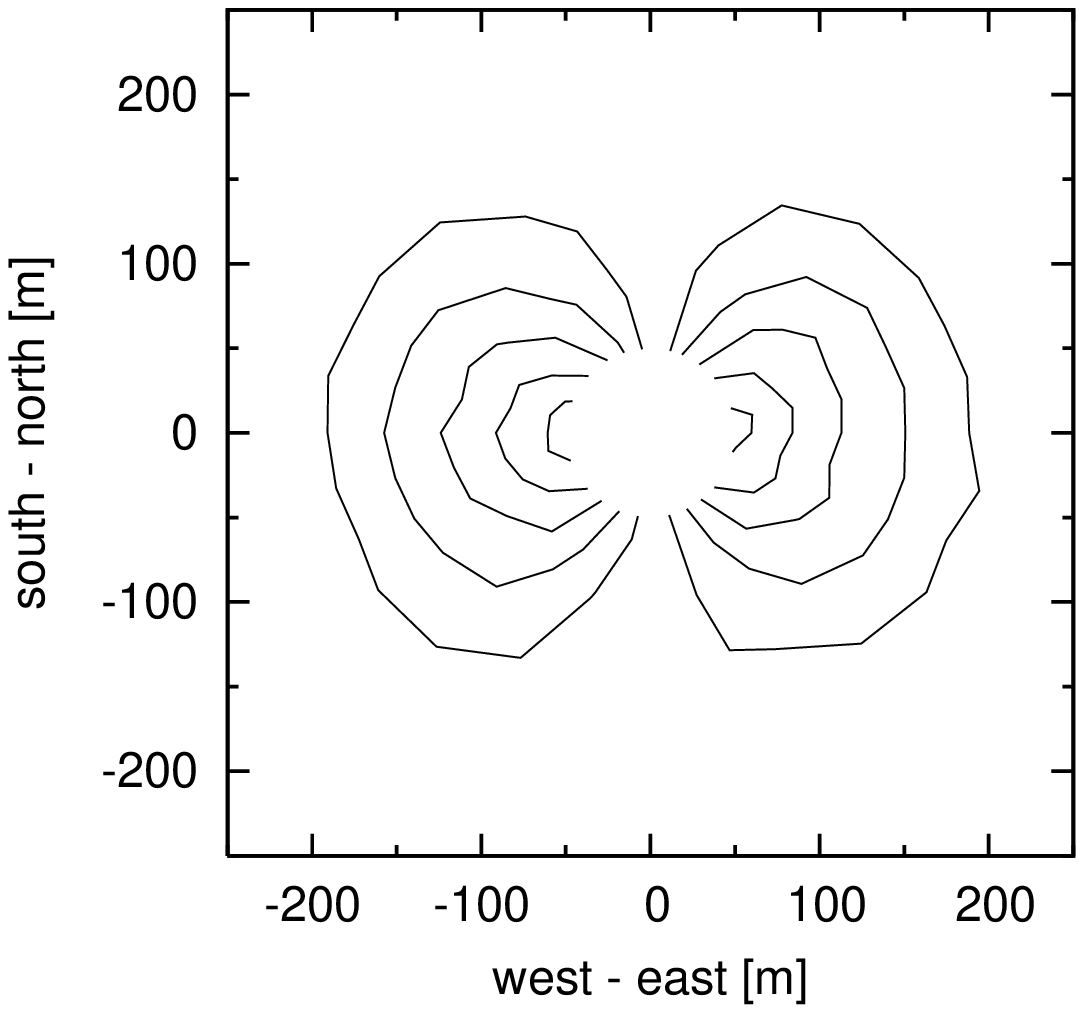}
		\end{minipage}
		\caption{Contour plots of the 60\,MHz field strength for emission from a $10^{17}$\,eV vertical air shower without any
		magnetic field.	 Contour levels are 0.03\,$\mu$Vm$^{-1}$MHz$^{-1}$ apart. The closest observer position to the shower core 
		is 50\,m.From left to right: total field strength, north-south and east-west polarisation component. The ``spike'' in 
		the lower-left part of the contours is associated with noise in the simulation.} \label{fig:contourB0}
	\end{figure}

\begin{figure}[h!]
		\begin{minipage}[h]{0.45\linewidth}
		\centering
		\includegraphics[width = 0.57\textwidth]{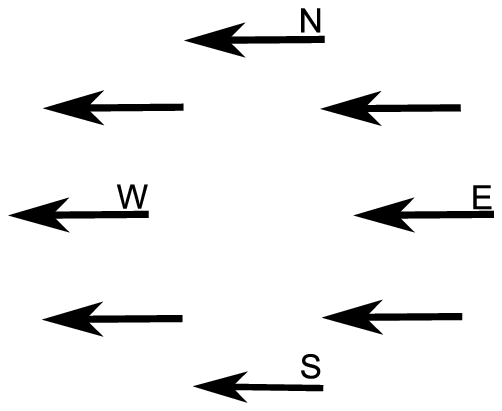}
		\end{minipage}
		\hspace{-0.3cm}
		\begin{minipage}[h]{0.45\linewidth}
		\centering
		\includegraphics[scale=0.6]{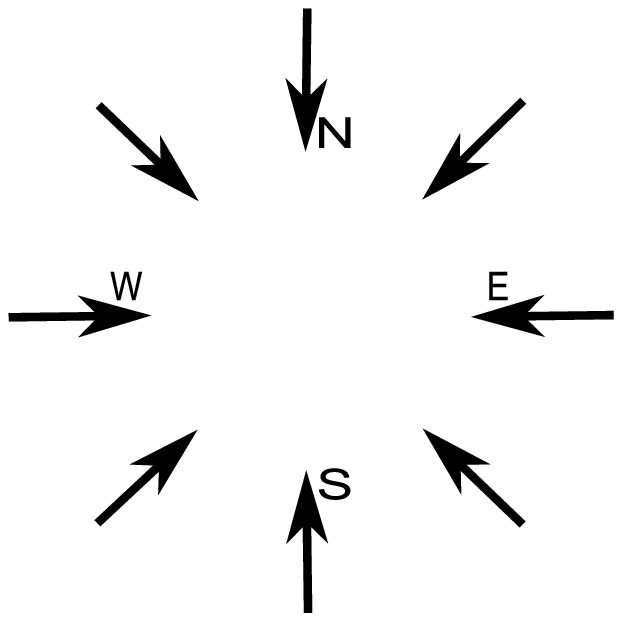}
		\end{minipage}
		\caption{Sketch of the polarisation vector. Left: uniform pattern as it is the case for pure geomagnetic emission.
		Right: radial pattern as it is the case for the net charge excess emission.}\label{fig:polarisations}
	\end{figure}

\begin{figure}[h!]	
		\begin{minipage}[b]{0.45\linewidth}
		\centering
		\begin{overpic}[angle = 270 , scale = 0.45]{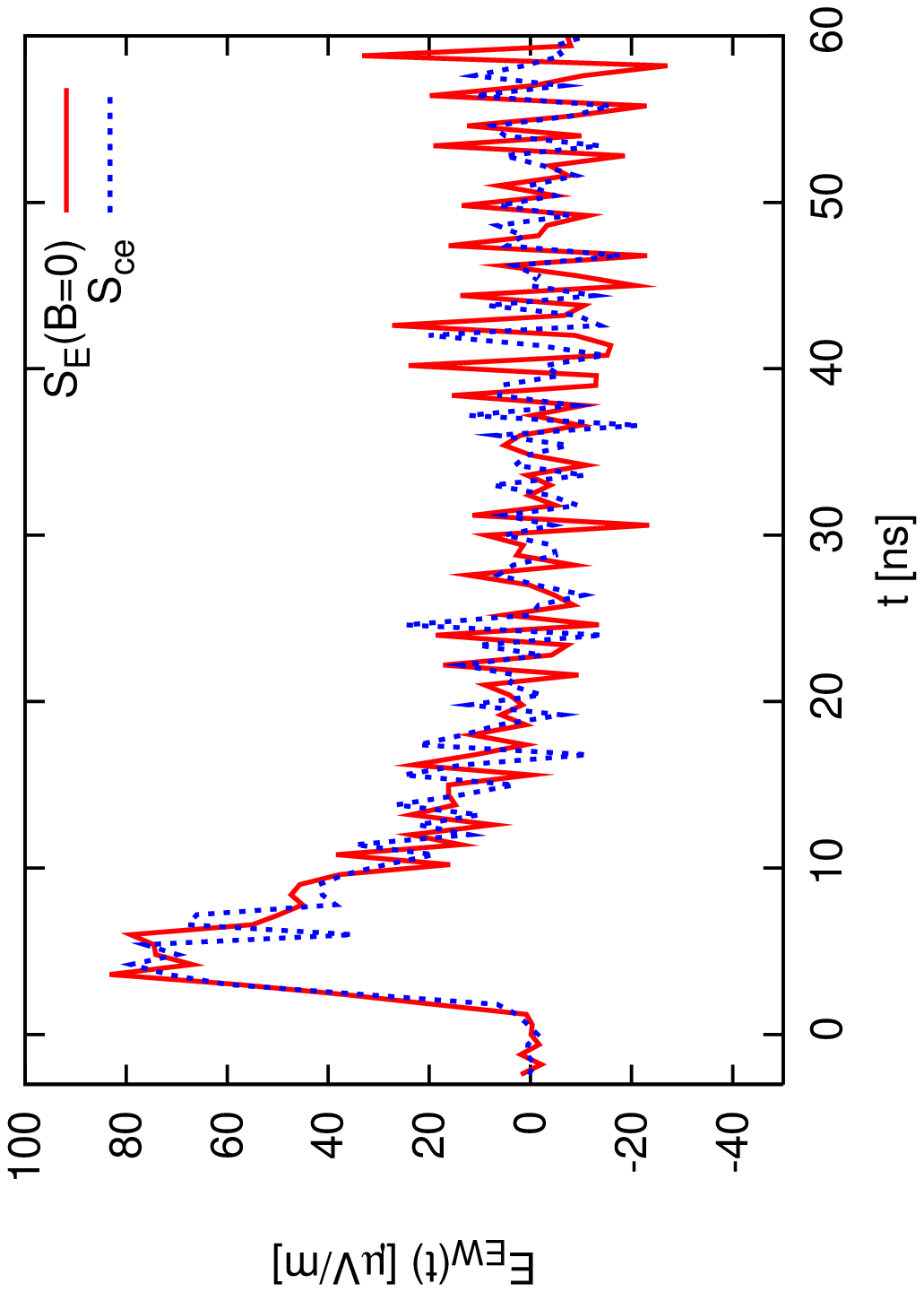}
		\put(42,60){\scriptsize{100\,m}}
		\end{overpic}
		\end{minipage}
		\hspace{0.6cm}
		\begin{minipage}[b]{0.45\linewidth}
		\centering
		\begin{overpic}[angle = 270 , scale = 0.45]{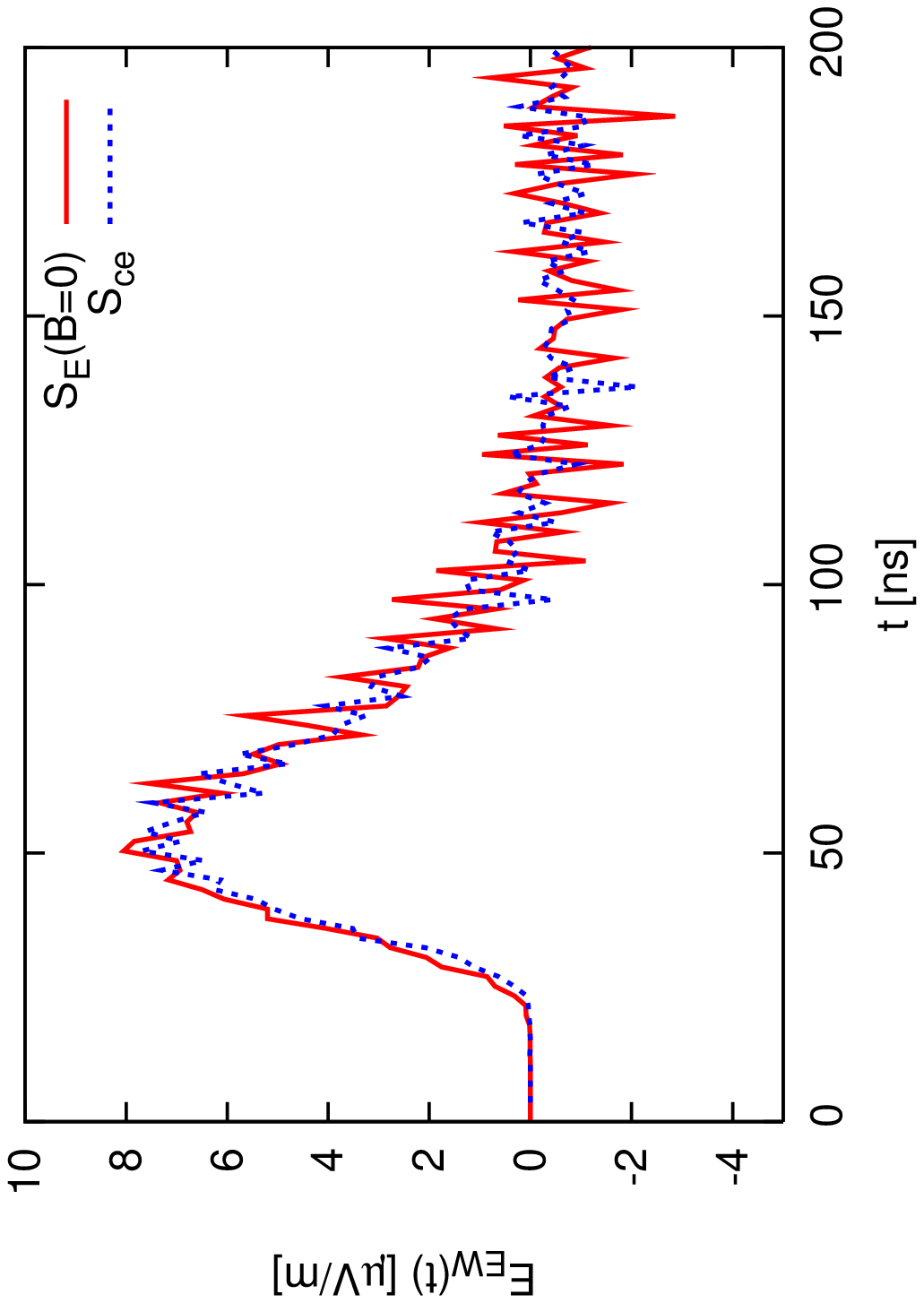}
		\put(42,60){\scriptsize{400\,m}}
		\end{overpic}
		\end{minipage}		
		\caption{Comparison of emission due to net charge excess without magnetic field (solid line) and the calculated signal for the net 
		charge excess (dashed line) from a shower with magnetic field. Displayed is in each case the east-west polarisation. Left: 
		100\,m distance from shower core. Right: 400\,m from shower core.}\label{fig:b0vsSum}
	\end{figure}

\begin{figure}[h!]	
		\centering
		\includegraphics[angle = 270 , scale = 0.45]{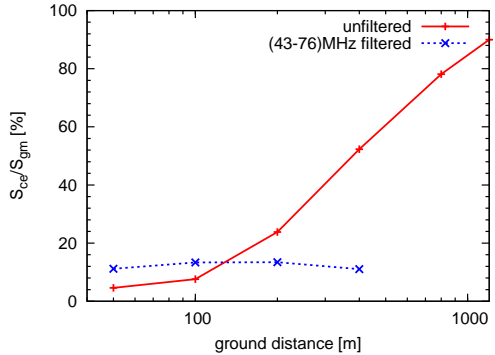}
		\hspace{0.6cm}
		\caption{Comparsion of the charge excess and the pure geomagnetic contribution on the radio signal of an vertical air 
		shower for the east-west polarisation component. The lines may not represent the correct interpolation between the points
		due to the logarithmic scale of the x-axis. The ratio of the filtered data is only shown for distances up to 400\,m, because
		then the signal is not anymore distinguishable from noise in this frequency range.}
	\label{fig:ratioCEvsVxB}
	\end{figure}

	The observed east-west asymmetry mentioned in section \ref{REAS2vsREAS3} arises from the fact that more electrons than 
	positrons exist in an extensive air shower \cite{Bergmann07}. This net charge excess of order 10-20\% leads to a contribution in the radio 
	signal even in the absence of any magnetic field.
	Hence, in this section an air shower was simulated with CORSIKA for the geometry and primary characteristics as already 
	mentioned at the beginning of this chapter, but in contrast to section \ref{REAS2vsREAS3}, the magnetic field strength was set
	to 0\,Gauss. Although the showers calculated with B\,=\,0.23\,Gauss and B\,=\,0\,Gauss do not represent the exact same particle 
	distributions, we can interpret the predictions for the shower with B\,=\,0\,Gauss as the contribution due to the net 
	charge excess of the shower used in section \ref{REAS2vsREAS3}. This allows us to test whether indeed the east-west
	asymmetry can be associated with the pure charge excess. In REAS2, radio emission is produced due to deflection of charged 
	particles in the magnetic field. Hence, there is no radiation without magnetic field. In contrast, REAS3 takes also emission 
	due to the variation of the net charge excess into account. A radially polarised component for emission due to charge excess is 
	expected, as seen also in the macroscopic approach \cite{WernerScholten2008a}. The radiation pattern for a shower with 
	B\,=\,0\,Gauss is indeed 
	radially polarised as illustrated in the contour plots of the 60\,MHz field strength in figure \ref{fig:contourB0}. For the 
	vertical component (not shown here) there is again no significant flux, as expected. Again, the closest observer position to the shower core is 50\,m. 
	Please note that the contour levels for the simulation without magnetic field are smaller than for the simulation with magnetic 
	field, i.e., the relative field strength of the charge excess emission at 60\,MHz is small in the distance range up to 200\,m.
	To study the influence of the net charge excess emission on the overall radio signal, it is helpful to look at the polarisation
	vectors in the plane perpendicular to the shower axis. For the pure geomagnetic emission, the polarisation vectors at all 
	observer positions point in the same direction as illustrated in the left sketch of figure \ref{fig:polarisations}. 
	The right sketch illustrates the polarisation vector for the emission due to the variation of the net charge excess. The 
	direction in which the vector points is changing with the observer position, following a radial pattern. Hence, for an 
	observer in the east the total signal 
	$S_E$ is given by 
	\begin{align}
	S_E = S_{gm} + S_{ce} \label{eq:s_e}
	\end{align}
	where $S_{gm}$ is the pure geomagnetic contribution and $S_{ce}$ the net charge excess contribution to the signal. For an 
	observer in the west, the total signal $S_W$ is composed of
	\begin{align}
	S_W = S_{gm} - S_{ce} \label{eq:s_w}
	\end{align}	
	With the signals measured east and west from the shower axis, the signal for the pure geomagnetic emission and for the net 
	charge excess can be calculated by
	\begin{align}
	S_{gm} = \frac{1}{2} (S_E + S_W) \qquad \text{and} \qquad S_{ce} = \frac{1}{2} (S_E - S_W). \label{eq:s_gm}
	\end{align}
	To verify the assumption made in equations (\ref{eq:s_e}) and (\ref{eq:s_w}), we calculated the signal of the charge excess as 
	it is resulting from above for the shower with B\,=\,0.23\,Gauss and compared this with the emission of the shower with\\
	B\,=\,0\,Gauss. Figure \ref{fig:b0vsSum} illustrates that both pulses match. Therefore, the east-west asymmetry in the azimuthal 
	emission pattern seen in the figures of section \ref{REAS2vsREAS3} is completely reducible to the emission of the net charge 
	excess in an air shower. \\
	Finally, it is interesting to quantify the relative strength of the charge excess emission with respect to the pure geomagnetic radio emission.
	Analyses studying the dependence of the radio signal on pure geomagnetic emission have already been done and have shown that there
	might be discrepancies between a pure geomagnetic model and the measured data \cite{Isar09}.
	The ratio of the net charge excess signal and the pure geomagnetic radiation can be calculated from equation 
	(\ref{eq:s_gm}) to quantify the relative influence of the net charge excess. Figure \ref{fig:ratioCEvsVxB} illustrates
	the ratio for the unfiltered full bandwidth amplitudes and the 43 to 76 MHz filtered 
	bandwidth amplitudes for the east-west polarisation component. For the filtered case the ratio in the plot is shown only for observers up to 400\,m lateral distance. 
	The reason is that the frequency spectra for a vertical shower drop fast with increasing lateral distance as was also seen 
	in figure \ref{fig:pulses}. Consequently, in the used frequency range the signal is not anymore distinguishable from 
	numerical noise at very large distances. For the amplitudes of the unfiltered pulses the charge excess has more and more influence
	with larger distances, from a few \% close to the core to around 90\% of the pure geomagnetic emission at 1200\,m. 
	However, for the filtered pulses the ratio is almost constant over the whole range, at a level of $\sim$10\%. 
	It is important to clarify that the emission of the net charge excess occurs due to the variation of the number of charged
	particles and not due to Cherenkov-like emission. Both processes have been described in the pioneering articles of Askaryan
	\cite{Askaryan1962}, \cite{Askaryan1965}, but the term ``Askaryan radiation'' is today generally interpreted as Cherenkov
	emission in dense media. For the inclusion of Cherenkov-like emission, a refractive index is needed which is not 
	``unity''. REAS3 approximates the index of refraction to be unity so far.

\section{Conclusion}\label{ch:conclusion}

	Using an ``end-point'' formalism, we have implemented a fully consistent Monte Carlo model for radio emission from air showers, which in particular
	includes also the emission contributions due to the variation of the number of moving charges.
	For the implementation into the simulation code	REAS, the description was 
	changed from a continuous treatment to a discrete description, because in the latter representation endpoint contributions can be added 
	canonically. REAS3 produces stable results. The pulse shape has changed from unipolar pulses in REAS2 to bipolar pulses in REAS3.
	This is also apparent in the frequency spectra of REAS3 simulations which drop to zero for small frequencies. Several results 
	illustrate the increased
	symmetry in the azimuthal emission pattern predicted by REAS3. Due to charge excess in extensive air showers it is evident that a small 
	asymmetry has to remain. This effect was shown by calculating the pure geomagnetic and charge excess contributions from the 
	asymmetries present in the full simulations and simulating radio emission in the complete absence of a magnetic field which 
	agree well within numerical uncertainties. In the absence of a magnetic field, REAS3 predicts a radially polarized 
	emission pattern. Due to the presence of this charge excess emission it is also obvious that the radio emission is not 
	purely of geomagnetic origin and thus cannot be described by a pure $\vec{v}\times \vec{B}$ dependence, 
	but that the signal polarisation depends on the exact observer position relative to the core. For a vertical shower, the relative strength of the 
	emission due to the variation of the net charge excess with respect to the pure geomagnetic emission increases from a few 
	\% close to the core up to $\sim$90\% at lateral distances of 1200\,m
	for the unlimited bandwidth pulse amplitudes. This net charge excess 
	affects the east-west symmetry as well as the polarisation of the radio emission. For the filtered 43\,MHz-76\,MHz bandwidth amplitudes the
	relative strength of the net charge excess with respect to the pure geomagnetic emission is constant at a level of approximately
	10\% for different lateral distances.\\
	Although the changes introduced with REAS3 are significant, many qualitative results obtained with earlier simulations are still valid (e.g.,
	approximately exponential lateral distributions, a dependence of the lateral slope on $X_{\mathrm{max}}$, the coherent scaling of the pulse
	amplitudes, and the fall-off of the frequency spectra to higher frequencies and larger distances). Certainly, the absolute amplitudes, pulse
	shapes and values of scaling parameters have changed with REAS3.
	With the implementation of emission due to the variation of the number of charged particles 
	in REAS3, for the first time a self-consistent time-domain model exists which takes the full complexity of air shower physics 
	as provided by CORSIKA simulations into account. The source code of REAS3 will be freely available on request. REAS3 therefore
	constitutes a widely usable radio simulation tool which can be employed to study radio emission from air showers initiated by
	arbitrary primary particles, using any of the available hadronic interaction models, realistic atmospheric profiles and arbitrary
	shower geometries (including near-horizontal showers, since the atmosphere in REAS3 is treated with a curved geometry).

\begin{ack}
The authors would like to thank S.\,Buitink, R.\,Engel, H.\,Falcke, O.\,Scholten and K.\,D.\,de Vries for very helpful discussions and
A.\,Haungs for very useful comments to the manuscript. This research has been supported by grant number VH-NG-413 of the Helmholtz Association.
\end{ack}







\bibliographystyle{model1-num-names}




\end{document}